\documentclass[12pt,twoside]{report}

\usepackage[utf8]{inputenc}
\usepackage{graphicx}
\graphicspath{ {figures/} }
\usepackage[a4paper,width=150mm,top=25mm,bottom=25mm,bindingoffset=6mm,headheight=15pt]{geometry}
\usepackage{palatino}
\usepackage{amsmath}
\usepackage{amssymb}
\usepackage[hidelinks]{hyperref} 
\hypersetup{
    colorlinks=true,
    linkcolor=blue,
    filecolor=magenta, 
    urlcolor=cyan,
}
\usepackage{tikz}
\usetikzlibrary{shapes,arrows,positioning,fit,backgrounds,calc}
\usepackage{xcolor}

\newcommand{\CYAN}[1]{\textcolor{cyan}{#1}}
\newcounter{mknot} % Notes from Mieke
\newenvironment{mknot}[1][]{\refstepcounter{mknot}\par\medskip
   \noindent \textbf{\CYAN{NotebyMieke~\themknot.}  #1} \rmfamily}{\medskip}

\newcommand{\RED}[1]{\textcolor{red}{#1}}
\newcounter{vcnot} % Notes from Vincenzo
\newenvironment{vcnot}[1][]{\refstepcounter{vcnot}\par\medskip
   \noindent \textbf{\RED{NotebyVincenzo~\thevcnot.}  #1} \rmfamily}{\medskip}

\usepackage{titlesec}
\titleformat{\chapter}[hang]{\fontsize{40}{50}\selectfont\bfseries}{\thechapter}{20pt}{\Huge\bfseries}
\titleformat{\paragraph}[hang]{\normalfont\normalsize\bfseries}{\theparagraph}{1em}{}
\titlespacing*{\paragraph}{0pt}{3.25ex plus 1ex minus .2ex}{1em}
\usepackage{fancyhdr}
\usepackage{parskip}
\usepackage[rightcaption]{sidecap}
\sidecaptionvpos{figure}{c} % Vertically centered captions
\usepackage{subcaption}
\usepackage[labelfont=bf,font=it]{caption}
\usepackage{csquotes}
\renewcommand{\mkbegdispquote}[2]{\itshape\openautoquote}

\usepackage{listings}
\usepackage{verbatim}
\lstdefinelanguage{json}{
    basicstyle=\ttfamily\small,
    numbers=none,
    keywordstyle=\color{blue},
    stringstyle=\color{green!50!black},
    commentstyle=\color{gray},
    frame=single,
    breaklines=true,
    showstringspaces=false,
    morestring=[b]",
    morestring=[b]',
    morecomment=[l]{//},
    morecomment=[s]{/*}{*/},
    identifierstyle=\color{black},
    keywords={false,true,null},
    ndkeywords={
        "id","name","state","openedDatasets","openedCases",
        "layerSettings","T1","colorMap","visible","tumor_mask",
        "opacity","currentScript"
    },
    ndkeywordstyle=\color{blue},
    sensitive=false
}
\lstdefinelanguage{JavaScript}{
  keywords={typeof, new, true, false, catch, function, return, null, catch, switch, var, if, in, while, do, else, case, break},
  keywordstyle=\color{blue}\bfseries,
  ndkeywords={class, export, boolean, throw, implements, import, this},
  ndkeywordstyle=\color{darkgray}\bfseries,
  identifierstyle=\color{black},
  sensitive=false,
  comment=[l]{//},
  morecomment=[s]{/*}{*/},
  commentstyle=\color{purple}\ttfamily,
  stringstyle=\color{red}\ttfamily,
  morestring=[b]',
  morestring=[b]"
}
\usepackage{dirtree}

\addtocontents{toc}{\protect\thispagestyle{empty}}

\begin{document}
    % Title page
    \pagestyle{empty}
    \begin{center}
    % University Logo and Name (included in the EPS file)
    \includegraphics[width=0.8\textwidth]{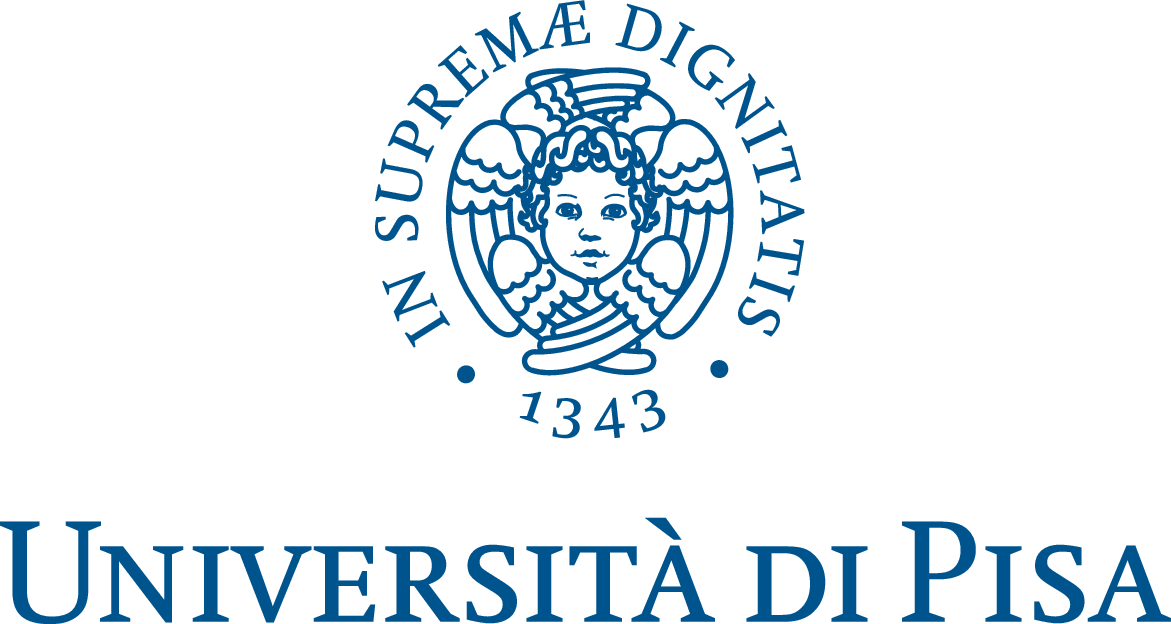}
    
    \vspace{0.2cm}
    {\Large\textsc{Computer Science Department}}
    
    \vspace{0.2cm}
    {\large\textsc{Master in Computer Science}}
    
    \vspace{5cm}
    {\Huge\bfseries VoxLogicA UI:}\\[0.3cm]
    {\huge\bfseries Supporting Declarative}\\[0.3cm]
    {\huge\bfseries Medical Image Analysis}
    
    \vfill
    
    \begin{minipage}[t]{0.45\textwidth}
        \begin{flushleft}
            \textbf{Candidate:}\\
            Antonio Strippoli
        \end{flushleft}
    \end{minipage}
    \hfill
    \begin{minipage}[t]{0.45\textwidth}
        \begin{flushright}
            \textbf{Supervisors:}\\
            Vincenzo Ciancia \\
            Fabio Gadducci \\
            Mieke Massink
        \end{flushright}
    \end{minipage}
\end{center} 
    
    \newpage \ \newpage
    
    % Quote
    \topskip0pt
\vspace*{\fill}
\begin{large}
\begin{center}
\textit{"Life is short, craft long, opportunity fleeting,\\ experimentations perilous, and judgment difficult."}
\end{center}
\begin{flushright}
\textbf{Hippocrates}
\end{flushright}
\end{large}
\vspace*{\fill}
    \newpage \ \newpage
    
    % Table of contents
    \tableofcontents

    \newpage \ \newpage
    
    % Main chapters
    \chapter{Introduction}
\setcounter{page}{1}
\pagestyle{fancy}

The role of spatial logic in medical image analysis has emerged as a promising approach through the VoxLogicA project, offering ways to describe and analyze anatomical structures in ways that align with domain-specific medical reasoning \cite{voxlogica2019, voxlogicaSpin2021}. Recent advancements, such as the application of spatial model checking for nevus segmentation, demonstrate its feasibility in achieving robust results comparable to expert annotations while maintaining computational efficiency and explainability \cite{belmonteBCLM21}.

This thesis work emerges from three complementary research initiatives:
\begin{itemize}
    \item \textbf{THE} (Tuscany Health Ecosystem) \cite{the2024} focuses on developing digital health ecosystems for a plethora of applications -- among which personalized healthcare and brain health. THE aims at preparation for technological transfer, and has a specific \emph{action} devoted to advancement of the technological nature of VoxLogicA;
    \item \textbf{STENDHAL} (Spatio-Temporal Enhancement of Neural nets for Deeply Hierarchical Automatised Logic), a PRIN project \cite{stendhal2023}, is a foundational and multi-disciplinary project devoted to enriching spatio-temporal analysis via neural-symbolic integration, combining neural networks with formal spatial model checking to enable expert-driven verification of medical protocols and safety properties in longitudinal healthcare studies;
    \item \textbf{FM4HD} (Formal Methods For the Healthcare Domain based on spatial information) \cite{fm4hd2022} focussed less on the inter-disciplinary issues of STENDHAL and more specifically on the challenges offered by the enrichment of the expressiveness of the spatio-temporal logics and its implementation.
\end{itemize}

Our work bridges a critical gap common to all three projects. Spatial model checking has shown theoretical promise through tools like VoxLogicA, and has demonstrated potential practical impact via a number of successful case studies. However, clinical adoption requires addressing fundamental human-computer interaction challenges. Thus, spatial model checking's clinical value can only be fully realized through \textit{simultaneous advances} in formal methods and interaction design. This creates a unique research tension: how can we preserve the formal guarantees of spatial logic while creating interfaces that align with both computer science researchers' and medical professionals' cognitive workflows?

\newpage

We want to demonstrate that, in the field of medical imaging, formal methods:

\begin{itemize}
    \item Can be operationalized through careful interface design;
    \item Enable new human-computer collaboration patterns in diagnostic workflows;
    \item Require usability studies to complete the research-to-practice cycle.
\end{itemize}

\section{Research Objectives}
This thesis addresses the need for a user-friendly interface to VoxLogicA, a spatial model checking tool for image analysis. The primary objectives include:

\begin{itemize}
    \item Conducting a comprehensive design study for a modern web interface for VoxLogicA, aimed at enhancing accessibility for medical professionals and researchers;
    \item Implementing the interface using modern web development technologies and best practices;
    \item Conducting empirical validation through structured user studies and real-world case analyses to assess interface efficacy.
\end{itemize}

\section{Thesis Structure}
This thesis is organized as follows:

Chapter 2 provides the multidisciplinary context of the research through four integrated pillars: spatial model checking's mathematical framework, neuroimaging analysis challenges, UI design principles for medical interfaces, and modern web architectures. The synthesis operationalizes theoretical insights through comparative technology assessments, justifying subsequent implementation choices.

Chapter 3 presents VoxLogicA UI as a modern web-based interface for spatial model checking in neuroimaging, designed to bridge the gap between VoxLogicA's advanced analytical capabilities and clinical/research usability requirements. Motivated by a critical analysis of legacy neuroimaging tools - revealing persistent usability challenges - the solution is designed and implemented using the latest web development technologies (e.g. Svelte and SkeletonUI) and best practices (e.g. MVVM architecture, RESTful API design, Docker deployment). The project architecture follows a layered pattern, with each layer having specific responsibilities and communicating only with adjacent layers.

Chapter 4 details the concrete realization of the VoxLogicA UI system through an integrated examination of implementation strategies across the full technology stack. The development environment lays the groundwork for the project, featuring docker containerization and continuous integration. The backend exposes a hierarchical security-conscious REST API structured around core medical imaging concepts (datasets, cases, layers), with a particular highlight being a novel workspace cloning mechanism addressing clinical collaboration needs through immutable snapshots that preserve data sovereignty while enabling derivative analyses. The frontend exposition progresses through a vertical slice of the visualization pipeline, from NiiVue WebGL integration to reactive layer management components, culminating in an analysis panel to write and run spatial logic queries. Finally, adaptive distribution implementations are discussed, accommodating heterogeneous clinical and research infrastructures through flexible distribution pathways spanning local containerized installations and cloud-native deployments.

Chapter 5 transitions from technical implementation to empirical validation via a multi-layered validation strategy combining rigorous usability testing and real-world exposure initiatives. The analyses were conducted through controlled user studies employing mixed-methods research design, combining quantitative metrics like System Usability Scale (SUS) scores with qualitative observational data. Two parallel studies - targeting computer science researchers and medical physics students - reveal divergent usability patterns that reflect the system's dual role as both research environment and clinical analysis tool. Public exposure of VoxLogicA UI is then described, ranging from engagement with the NiiVue team for future technical collaborations to clinical engagement efforts at Bari Hospital and Bari University with a radiologist, a general practitioner and some students.

Chapter 6 synthesizes the project's contributions through a critical reflection on achieved outcomes and remaining challenges.

    % Background chapters
    \chapter{Background Knowledge}

\section{Formal Methods}

In the context of modern software systems, where complexity continues to grow and reliability becomes increasingly critical, formal methods emerge as a fundamental approach to ensure system correctness \cite{formalMethodsBook2022, nasaFormalMethods2002}. At their core, formal methods represent the application of mathematics to the modelling and analysis of Information and Communication Technology (ICT) systems, providing a \textbf{rigorous framework for verification and validation}.

Formal methods enable developers and researchers to verify that a model of a system design or implementation satisfies properties derived from its specification. When a model fails to meet any of these properties, a defect is identified; conversely, a model is deemed correct when it fulfils all the properties of interest.

Formal methods enable the early integration of verification in the design process, providing systematic verification techniques, and significantly reducing the time required for system validation. Their importance is particularly evident in safety-critical systems, where they are considered a \textit{highly recommended} verification approach. Notable examples from industry underscore their value: formal verification techniques could have prevented well-known failures in systems such as the \textbf{Ariane-5 rocket}, \textbf{Mars Pathfinder}, and \textbf{Intel's Pentium II processor}.

\begin{figure}[h]
    \centering
    \includegraphics[width=\textwidth]{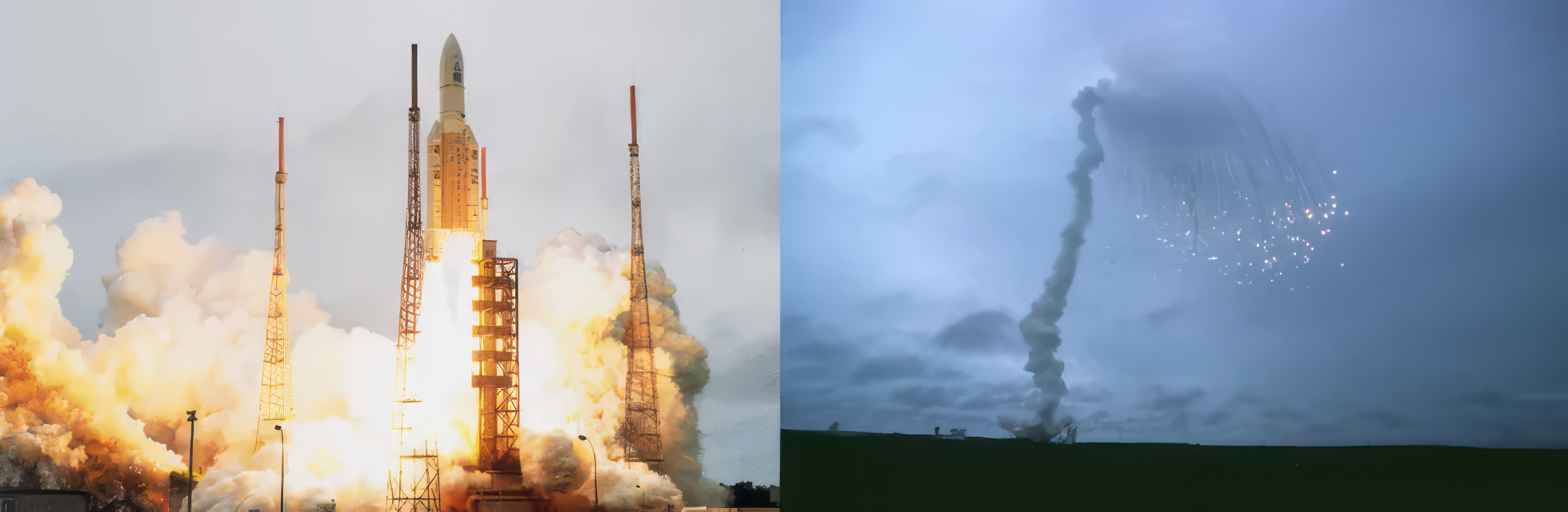}
    \caption{Ariane-5 rocket's catastrophic failure during its maiden flight in 1996, costing over \$370 million. The disaster was caused by a software error in the Inertial Reference System (SRI), which led to an operand error and subsequent system shutdown. Formal methods could have prevented this failure by identifying and addressing the software design flaws and exception handling issues before launch. \cite{ariane5Flight501}}
    \label{fig:ariane5}
\end{figure}

\subsection{Model Checking: an overview}
Model-based verification techniques employ mathematical models to precisely and unambiguously describe potential system behaviors. This rigorous approach often uncovers incompleteness, ambiguities, and inconsistencies inherent in informal system specifications. Among these techniques, model checking stands out as a formal verification method that \textbf{systematically inspects all possible states of a system model} to ascertain compliance with a given set of properties \cite{principlesModelChecking}. The appeal of model checking lies in its complete automation and its ability to provide counterexamples when a model fails to meet a property, offering invaluable insights for debugging.

Typical properties that can be verified using model checking are of a qualitative nature, such as whether the generated result is OK, or whether the system can reach a deadlock situation. Timing properties can be taken into account, such as whether a deadlock can occur within one hour after a system reset.

Relevant to this work, spatial properties can also be verified through model checking. For instance, spatial logic can be applied to ensure that a detected tumor region does not exceed certain boundaries or to assess the spatial relationship between different anatomical structures.

In the following, we provide a brief theoretical introduction to model checking, discussing system modelling, property specification, and checking process.

\subsubsection{System Modelling}

System modelling involves the \textbf{abstraction of a real-world system into a mathematical representation}. This representation is typically achieved through the use of finite-state automata (FSA), which are mathematical models composed of a finite number of states and transitions between those states.

\paragraph{Definition: Finite-State Automaton (FSA)}
A finite-state automaton is defined as a tuple \( A = (S, S_0, \Sigma, \delta, F) \), where:
\begin{itemize}
    \item \( S \) is a finite set of states;
    \item \( S_0 \subseteq S \) is a set of initial states;
    \item \( \Sigma \) is a finite set of input symbols, known as the alphabet;
    \item \( \delta: S \times \Sigma \rightarrow S \) is the transition function;
    \item \( F \subseteq S \) is a set of accepting states.
\end{itemize}

The automaton begins in one of the initial states and evolves through states according to the transition function \( \delta \), based on the input symbols from the alphabet \( \Sigma \).

\paragraph{State and Transition}
\begin{itemize}
    \item \textbf{State:} a state in an FSA represents a unique configuration of the system at a given point in time. It encapsulates all necessary information to determine the system's future behavior;
    \item \textbf{Transition:} a transition is a directed edge between two states, indicating the system's evolution from one state to another in response to an input symbol.
\end{itemize}

\subsubsection{Property Specification}
Properties to be checked are described using a property specification language.

While this work is focused on spatial properties, we will briefly introduce \textbf{temporal logic} to give more context to the reader.

\paragraph{Linear Temporal Logic (LTL) and Computation Tree Logic (CTL)}
Temporal logics are formal languages used to specify system properties over time. The two most common variants are:

\begin{itemize}
    \item \textbf{Linear Temporal Logic (LTL)} views time as a linear sequence of states, expressing properties about paths in a computation. It includes temporal operators such as:
    \begin{itemize}
        \item \( \bigcirc \) (next): property holds in the next state;
        \item \( \square \) (always): property holds in all future states;
        \item \( \lozenge \) (eventually): property holds in some future state;
        \item \( \mathcal{U} \) (until): first property holds until second property becomes true.
    \end{itemize}
    
    \item \textbf{Computation Tree Logic (CTL)} views time as a branching tree of possible futures, adding path quantifiers:
    \begin{itemize}
        \item \( \forall \) (for all paths);
        \item \( \exists \) (exists a path).
    \end{itemize}
\end{itemize}

\subsubsection{Model Checking Process}
The model checking process is usually divided into three main phases:

\begin{itemize}
    \item \textbf{Modelling Phase:} this involves creating a model of the system and formalising the properties to be verified;
    \item \textbf{Running Phase:} in this phase, the model checker algorithmically checks the validity of the property in all the states of the system model;
    \item \textbf{Analysis Phase:} this stage involves analysing the results. If a property is violated, a counterexample is generated, which is then used to refine the model, design or property.
\end{itemize}

\subsection{Spatial Model Checking}
Spatial model checking extends traditional model checking by incorporating spatial dimensions into the analysis of systems \cite{spatialLogicWithTimeAndQuantifiers}. This approach is essential for systems distributed across physical spaces, where \textit{locality} and the spatial distribution of objects can significantly influence system behavior and performance. Understanding these spatial properties is crucial for accurately modeling and verifying systems such as \textit{bike-sharing networks} \cite{CLMPV16, lmcs:3774}, \textit{cyber-physical systems} \cite{Ciancia2018}, and \textit{medical imaging applications} \cite{belmonteBCLM21, hybridAIvoxlogica2024, voxlogica2019, voxlogicaSpin2021, spatialLogicsClosureSpaces}.

Dating back to early logicians such as Tarski, \textit{spatial logics} arise as formalisms for describing \textit{geometrical entities} and \textit{configurations}, and are well-developed in terms of descriptive languages and computability/complexity aspects. These logics serve as invaluable tools for analyzing systems with spatial dimensions.

In the following, we will provide a brief overview of the most important concepts in topology, before delving into the specifics of spatial logics.

\subsubsection{Topology}

Most of the following content is based on \cite{spatialLogicWithTimeAndQuantifiers}.

% Thinking about what Fabio wrote, I think it makes sense to remove epsilon-balls and open/closed sets, as they are not used in the rest of the chapter. Keeping them for now to wait for your feedback.

% \paragraph{The \(\epsilon\)-ball}
% Given a point \( x \) in \(\mathbb{R}^n\), the \(\epsilon\)-ball around \( x \) is defined as:
% $$B_{\epsilon}(x) = \{y \in \mathbb{R}^n : d(x,y) < \epsilon \}$$
% where \( d(x,y) \) is the usual Euclidean distance.

% \paragraph{Open set}
% A set \( U \subseteq \mathbb{R}^n \) is called \textbf{open} if for every \( x \in U \) there is an \( \epsilon > 0 \) such that \( B_{\epsilon}(x) \subseteq U \).

% \paragraph{Closed set}
% In mathematics, specifically in the context of topology, a closed set is a subset of a given topological space that contains all its limit points.

\paragraph{Space}
A \textbf{space} \( \mathcal{C} \) is a pair \( (X,C) \) where \( X \) is a set, and the \textbf{closure operator} \( C : 2^X \rightarrow 2^X \) assigns to each subset of \( X \) its closure, obeying to the following laws, for all \( A,B \subseteq X \):
\begin{itemize}
    \item \textbf{Closure of the empty set}: \( C(\emptyset) = \emptyset \);
    \item \textbf{Preservation of finite unions}: \( C(A \cup B) = C(A) \cup C(B) \).
\end{itemize}

\subparagraph{Complement and Interior}
Given \( (X, C) \) a space, and \( A \subseteq X \), then:
\begin{itemize}
    \item \textbf{Complement}: the complement of \( A \) in \( X \) is defined as \( A^c = X \setminus A \);
    \item \textbf{Interior operator}: the interior operator is defined as \( I(A) = C(A^c)^c = X \setminus C(S \setminus X) \), and it is also referred as the \textit{dual operator} of the closure.
\end{itemize}

\paragraph{Types of Spaces}
Given \( (X, C) \) a space, it is:
\begin{itemize}
    \item \textbf{Complete}: if \( C(\bigvee_{i\in I} X_i) = \bigvee_{i\in I} C(X_i) \) for any \( I \);
    \item \textbf{Pre-topological}: if \textit{extensivity} - i.e. \( X \subseteq C(X) \) - holds;
%    \item \textbf{Alexandrov}: if it is \textit{pre-topological} and \textit{complete};
    \item \textbf{Topological}: if it is \textit{pre-topological} and the closure operator is  \textit{idempotent} - i.e. \( C(C(X)) = C(X) \) - holds.
\end{itemize}

Note that if $X$ is finite then the space is always complete.

\subsubsection{Relations}
Given a set \( S \), a relation \( R \) is a function \( R : S \rightarrow 2^S \).

Relations are tightly related with simple graphs and unlabelled Kripke frames, which is what we are interested in in this work.

\paragraph{Lifting}
Given a set \( S \) and a relation \( R \), \( 2^R : 2^S \rightarrow 2^S \) is the lifting of \( R \), defined as: \( 2^R(X) = \bigcup_{x \in X} R(x) \).

\paragraph{Induced Relation}
Given a space \( \mathcal{C} = (S,C) \), the induced relation from \( \mathcal{C} \) is a function \( R_\mathcal{C} : S \rightarrow 2^S \) defined as \( R_\mathcal{C}(x) = C({x}) \).

\paragraph{Induced Complete Space}
Given a relation \( R : S \rightarrow 2^S \), the induced complete space from \( R \) is \( \mathcal{C}_R = (S, C_R) \), defined as \( C_R(X) = 2^R(X) \).

\paragraph{Logics interpretation on complete spaces and relations}
The following 2 points are true:
1. Let \( R : S \rightarrow 2^S \) be a relation. Then \( R_{\mathcal{C}_R}(x) = R(x) \) for all \( x \in S \). In other words, the induced relation obtained by the induced complete space from \( R \) is \textbf{equivalent} to \( R \);
2. Let \( \mathcal{C} \) be a complete space. Then \( \mathcal{C}_{R_\mathcal{C}}(X) = R(X) \) for all \( X \subseteq S \). In other words, the induced complete space obtained by the induced relation from \( \mathcal{C} \) is \textbf{equivalent} to \( \mathcal{C} \).

This leads to an \textbf{important point}: interpreting logics on \textbf{complete spaces} is \textbf{the same} as using as models the underlying \textbf{relations}. Also, axioms holding for complete spaces (see above) turn out to be state topological properties of such relations. By direct consequence, this allows us to treat graphs and topological spaces uniformly.

\newpage

\subsubsection{Spatial Logics}
Spatial logics extend traditional logical frameworks by incorporating spatial operators and relations. These logics provide formal languages for expressing and reasoning about:

\begin{itemize}
    \item \textbf{Spatial Properties}: including location, distance, topology, and morphology;
    \item \textbf{Spatial Relations}: such as adjacency, containment, and connectivity;
    \item \textbf{Spatial Operators}: near, touch, inside, outside, etc.
\end{itemize}

\paragraph{Spatial Logic for Closure Spaces (SLCS)}
SLCS is a modal logic specifically designed for reasoning about spatial properties in closure spaces \cite{spatialLogicsClosureSpaces}. It extends classical propositional logic with spatial operators that capture topological concepts. The syntax of SLCS is defined by the following grammar:

\[ \Phi ::= \texttt{true} \mid a \mid \lnot \Phi \mid \Phi \land \Phi \mid \overrightarrow{\rho} \Phi [\Phi] \mid \overleftarrow{\rho} \Phi [\Phi] \]

where:
\begin{itemize}
    \item \( a \) ranges over a set of atomic propositions;
    \item \( \overrightarrow{\rho} \) is the "forward reachability" operator;
    \item \( \overleftarrow{\rho} \) is the "backward reachability" operator.
\end{itemize}

The semantics of SLCS is defined over closure models \( \mathcal{M} = ((X,C), \mathcal{V}) \), where:
\begin{itemize}
    \item \( (X,C) \) is a closure space;
    \item \( \mathcal{V} \) is a valuation function mapping atomic propositions to subsets of \( X \).
\end{itemize}

Concerning the semantics of SLCS, the two most important operators are:
\begin{itemize}
    \item \( s \models \overrightarrow{\rho} \Phi_1 [\Phi_2] \) if there exists a spatial path \( ss_1 \ldots s_n \) in \( \mathcal{T} \) such that \( s_n \models \Phi_1 \) and \( s_j \models \Phi_2 \) for all \( j = 1, \ldots, n-1 \);
    \item \( s \models \overleftarrow{\rho} \Phi_1 [\Phi_2] \) if there exists a spatial path \( s_0 \ldots s_{n-1}s \) in \( \mathcal{T} \) such that \( s_0 \models \Phi_1 \) and \( s_j \models \Phi_2 \) for all \( j = 1, \ldots, n-1 \).
\end{itemize}

A practical application of SLCS has been demonstrated in medical image analysis through the VoxLogicA project, which is discussed in detail in \ref{sec:voxlogica}.

% \mieke{At this point I would provide a very simple example that illustrates spatial model checking and how images are represented as models. Something along the lines of the example in this video by Giovanna could be suitable: https://www.youtube.com/watch?v=yn705eavNfU  (at time 1:21), perhaps with a few more properties illustrated. The idea of the example is to bridge the gap between the very formal definitions above and the exploitation/transformation of the theory into actual spatial model checking. Maybe we can also add a few derived operators such as near and grow. For self-containedness, it might be an idea to also include the brain tumour segmentation specification that is also used in the poster and at https://vincenzoml.github.io/VoxLogicA/ adding some explanatory text. Of course, for details one can refer to the relevant publications themselves. The idea is to provide a clear insight of the context for which the GUI is developed.} % ANTONIO: the image from the poster was already included in this thesis and I believe serves as a perfect example already: I referred to the VoxLogicA background chapter here, instead of adding another example.

    \newpage
\section{Medical Image Analysis}
Medical image analysis is a rapidly evolving field at the intersection of medicine, computer science, and engineering. It involves the development and application of computational techniques to interpret and analyze medical images, which are crucial for diagnosing diseases, planning treatments, and conducting medical research \cite{medicalImageAnalysisDL}.

The primary goal of medical image analysis is to extract from images meaningful information that can aid in clinical decision-making. This process often involves several steps, including image acquisition, preprocessing, segmentation, feature extraction, and classification. Each of these steps requires sophisticated algorithms and techniques to ensure accuracy and reliability. For instance, image segmentation, which involves partitioning an image into meaningful regions, is a critical step that can significantly impact the subsequent analysis and interpretation.

\subsection{Neuroimaging}
This work focuses on a specific domain within medical image analysis: \textbf{neuroimaging}, with a particular emphasis on \textbf{brain tumor analysis}. Neuroimaging is a specialized field that involves the use of imaging techniques to study the structure and function of the brain \cite{VillanuevaMeyer2017}.

Brain tumors represent a significant area of study within neuroimaging due to their complex nature and the critical need for accurate diagnosis and treatment planning. Advanced imaging techniques are essential for identifying tumor boundaries, assessing tumor progression, and evaluating treatment responses. By leveraging these techniques, researchers and clinicians can develop more effective, targeted therapies that improve patient outcomes.

Next, we will briefly introduce the \textbf{Brain Tumor Segmentation (BraTS) dataset} and discuss the current approaches and tools used for brain tumor segmentation.

\subsubsection{NIfTI Format (.nii)}

The \textbf{Neuroimaging Informatics Technology Initiative (NIfTI)} format is a widely adopted standard for storing volumetric neuroimaging data. Files in this format typically have the extension \texttt{.nii} and are designed to efficiently handle complex imaging data \cite{niftiFileFormat}.

\textit{NIfTI files} can encapsulate a range of neuroimaging data types, including:
\begin{itemize}
    \item \textbf{3D Structural Images:} these include high-resolution scans such as T1-weighted and T2-weighted MRI, which are essential for anatomical studies and clinical diagnostics;
    \item \textbf{4D Functional Images:} this category encompasses functional MRI (fMRI) time series, which are crucial for understanding brain activity over time.
\end{itemize}

\newpage

The image data stored in NIfTI files can represent various types of brain scans, each serving distinct purposes in research and clinical practice:

\begin{itemize}
    \item \textbf{T1-weighted MRI:} known for its high-resolution anatomical detail, T1-weighted imaging is fundamental in structural brain analysis;
    \item \textbf{T2-weighted MRI:} useful for identifying lesions and abnormalities in brain tissue;
    \item \textbf{Diffusion Tensor Imaging (DTI):} provides insights into the white matter tracts of the brain, aiding in the study of connectivity and integrity;
    \item \textbf{Functional MRI (fMRI):} captures dynamic brain activity, allowing researchers to map functional areas and networks;
    \item \textbf{Fluid Attenuated Inversion Recovery (FLAIR) MRI:} useful for identifying lesions in the brain, such as those caused by multiple sclerosis or tumors;
    \item \textbf{Positron Emission Tomography (PET) Scans:} offer metabolic and molecular information, complementing structural/functional imaging.
\end{itemize}

\subsubsection{Brain Tumor Segmentation Challenge (BraTS)}
The \textbf{Brain Tumor Segmentation (BraTS) dataset} is a pivotal resource in the field of medical image analysis, specifically tailored for the segmentation of brain tumors. It serves as a benchmark for evaluating and comparing the performance of various segmentation algorithms.

BraTS is renowned for its inclusion of multi-institutional, multi-parametric magnetic resonance imaging (mpMRI) scans, which encompass four different MRI modalities: T1-weighted, T1-weighted post-contrast, T2-weighted, and Fluid Attenuated Inversion Recovery (FLAIR). These modalities are crucial for capturing the diverse characteristics of brain tumors, including their heterogeneity and infiltration into surrounding tissues. The dataset is annotated with expert delineations of tumor sub-regions, such as the enhancing tumor, the peritumoral edema, and the necrotic core, providing a rich ground truth for algorithm development.

Moreover, the BraTS challenge, an annual competition associated with the dataset, encourages the development of innovative segmentation techniques. Participants submit their segmentation results, which are then evaluated using standardized metrics such as the \textbf{Dice Similarity Coefficient}, sensitivity, and specificity \cite{bratsWebsite}.

\begin{figure}[ht]
    \centering
    \includegraphics[width=0.55\textwidth]{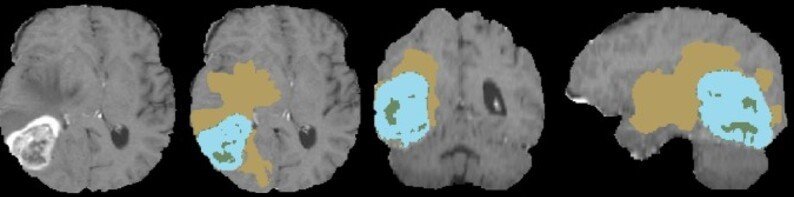}
    \caption{Axial slices of a brain MRI scan from the BraTS dataset showing tumor segmentation overlaid on T1-weighted images. The segmentation shows different tumor sub-regions: the peritumoral edema (yellow/beige), the necrotic and non-enhancing tumor core (light blue), and the enhancing tumor (dark green) \cite{brats2021Kaggle}.}
    \label{fig:brats_dataset_cover_2021}
\end{figure}

\subsubsection{Dice Similarity Coefficient (DSC)}
The \textbf{Dice Similarity Coefficient (DSC)} is a statistical tool used to gauge the similarity between two sets \cite{diceSimilarityCoefficient}. It is particularly relevant to this work as it provides a quantitative measure of overlap between the predicted segmentation and the ground truth. The DSC is computed as follows:

\[
\text{DSC} = \frac{2 \times |X \cap Y|}{|X| + |Y|}
\]

where \(X\) is the set of predicted segmentation voxels\footnote{A voxel (\textbf{vo}lumetric pi\textbf{xel}) represents a value on a regular grid in three-dimensional space, serving as the fundamental unit of volumetric medical imaging data.} and \(Y\) is the set of ground truth voxels. A DSC of 1 indicates perfect agreement, while a DSC of 0 indicates no overlap.

The \textbf{ground truth} is typically obtained as an \textbf{average of three manual expert segmentations}, acknowledging that inter-expert segmentations may vary. This variability explains why a perfect match between prediction and ground truth is usually not found, and a 90\% overlap is already considered an excellent result.

\subsubsection{Current Approaches and Tools}
\textbf{Traditional image processing techniques} dominated medical image analysis until the early 2000s, relying on mathematical and statistical methods to extract meaningful information \cite{sahiner2018deep}. These approaches included edge detection, morphological operations, and intensity-based segmentation, laying the foundation for automated image analysis \cite{unesp2023}. Watershed algorithms, active contours, and statistical shape models were particularly influential in addressing challenges such as organ segmentation and anatomical structure detection. However, these methods often struggled with the complexity and variability of medical images, requiring extensive manual parameter tuning and failing to generalize across different imaging modalities or patient populations \cite{despotovic2015mri}.

The evolution from traditional machine learning to \textbf{deep learning} approaches marked a paradigm shift in medical image analysis \cite{sahiner2018deep}. Early machine learning methods relied on handcrafted features (such as HOG, SIFT, and LBP) combined with classical algorithms like SVMs and Random Forests. The advent of deep learning architectures, particularly \textbf{Convolutional Neural Networks (CNNs)}, has eliminated the need for manual feature engineering. These deep architectures, exemplified by the widely-adopted \textbf{U-Net} for segmentation tasks, enable automatic discovery of optimal representations directly from raw image data.

More recently, \textbf{spatial logic} approaches have gained attention as a means to incorporate domain knowledge and anatomical constraints into image analysis frameworks. This approach leverages logical expressions to define spatial relationships and properties within medical images, enhancing the interpretability of image analysis tasks.

Spatial logic is particularly beneficial in scenarios where traditional image processing techniques struggle, such as in the presence of complex anatomical \\ structures or when precise localization of features is required. By embedding spatial constraints directly into the analysis process, spatial logic can improve the robustness and reliability of segmentation and classification tasks.

Moreover, \textbf{spatial model checking} is unique in its ability to visualize the results, a feature not commonly found in traditional model checking. This capability has facilitated its application in the medical field, despite requiring special arrangements to accommodate the complexity and specificity of medical data.

\paragraph{VoxLogicA}
\label{sec:voxlogica}
\textbf{VoxLogicA} is a sophisticated \textbf{spatial model checker} designed for \textit{declarative image analysis}, particularly in the domain of medical imaging \cite{voxlogica2019}. It leverages \textbf{spatial and spatio-temporal model checking techniques} to facilitate the semi-automated contouring of medical images, such as those used in brain tumor segmentation.

The core innovation of VoxLogicA lies in its use of \textbf{spatial logic}, which allows for the specification of spatial properties and relationships within images. This approach provides a high level of \textit{explainability} and \textit{replicability}, which are critical in medical imaging applications. The tool's logical framework is based on the \textbf{Spatial Logic for Closure Spaces (SLCS)}, covered in the previous sections.

The \texttt{ImgQL} language, which forms the basis of VoxLogicA's specification framework, allows users to define complex spatial queries and constraints. This language supports logical operators for \textit{reachability} and \textit{statistical similarity}, critical for tumor segmentation and similar tasks. The tool's ability to incorporate \textit{domain-specific knowledge} through logical expressions enhances output interpretability, making it a valuable asset for medical professionals.

VoxLogicA's execution engine is implemented in \texttt{FSharp}, utilizing the \texttt{.NET Core} framework to ensure cross-platform compatibility and efficient execution. It also integrates the \textbf{Insight Segmentation and Registration Toolkit (ITK)} library with a declarative specification language, enabling rapid development and execution of image analysis tasks.

Remarkably, the tool achieves significant performance improvements over previous model checkers, such as \texttt{topochecker}, through optimized algorithms and parallel execution capabilities. This efficiency enables \textit{interactive development and analysis} of 3D medical images, a crucial feature for clinical applications.

\begin{figure}[ht]
    \centering
    \includegraphics[width=0.68\textwidth]{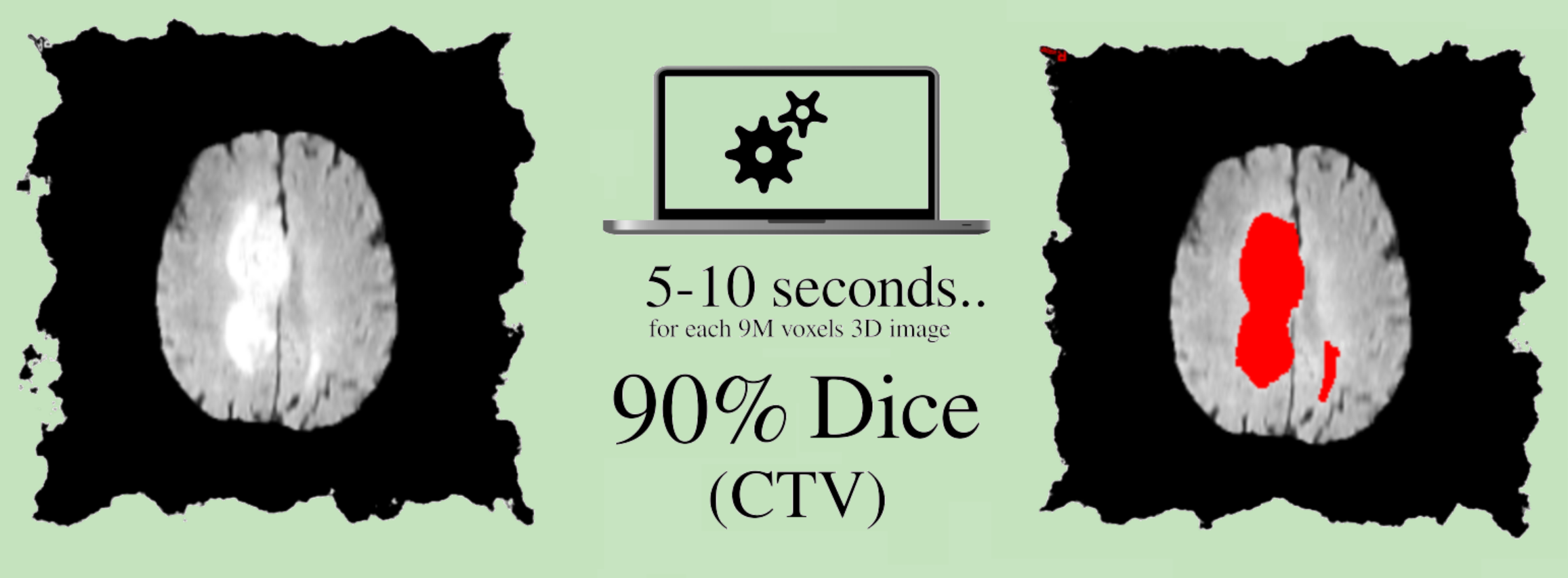}
    \caption{Example of brain tumor segmentation using VoxLogicA, achieving 90\% Dice coefficient in 5-10 seconds for each 3D image containing 9 million data points. For more information, please check the \href{https://vincenzoml.github.io/VoxLogicA/}{VoxLogicA website}.}
    \label{fig:voxlogica_example}
\end{figure}
    \section{Modern Web Development}
In the rapidly evolving landscape of web development, the quest for more efficient, scalable, and user-friendly applications has led to the adoption of innovative paradigms and frameworks. This chapter examines contemporary methodologies shaping web development, with particular focus on the tools and techniques employed in this thesis: reactive programming paradigms, the Model-View-ViewModel (MVVM) architectural pattern, and the Svelte 5 framework.

\subsection{Reactive Programming Paradigms}
\textbf{Reactive programming} is a declarative programming paradigm concerned with data streams and change propagation \cite{reactiveProgrammingWikipedia, reactiveProgrammingTechTarget}. This approach enables developers to concisely express dynamic application behavior through:

\begin{itemize}
    \item Treatment of data as \textbf{continuous streams} (time-ordered sequences of related event messages);
    \item Real-time \textbf{response mechanisms} using observer and handler functions;
    \item Automatic \textbf{change propagation} through the system, updating dependent values without explicit reassignment.
\end{itemize}

For example, in reactive programming, the declaration \texttt{a = b + c} ensures automatic updates to \texttt{a} whenever \texttt{b} or \texttt{c} changes, without requiring explicit re-execution.

The paradigm is particularly useful and widespread in \textbf{web development}, where the user interface is often dynamic and requires real-time updates.

\subsection{Architectural Pattern: Model-View-ViewModel (MVVM)}
The \textbf{Model-View-ViewModel (MVVM)} architectural pattern decouples application user interfaces from business logic \cite{builtinMVVM}, structuring applications through three interconnected components:

\begin{itemize}
    \item \textbf{Model}: represents the data and business logic of the application;
    \item \textbf{View}: represents the user interface;
    \item \textbf{ViewModel}: acts as an intermediary between the Model and the View, handling data binding and user interface updates.
\end{itemize}

It is worth noting that implementation instructions for MVVM are not standardized, and it has been a topic of debate in the software engineering community for a long time now \cite{russellEastMVVM}. Thus, in this work we will show just one of the many possible implementations of MVVM.

\newpage

\subsection{TypeScript}
\textbf{TypeScript} is a strongly typed programming language that builds on JavaScript by adding static type definitions \cite{typescriptHandbook}. It is a superset of JavaScript, meaning that any valid JavaScript code is also valid TypeScript code. The key features of TypeScript include:

\begin{itemize}
    \item \textbf{Type System}: combines structural typing with gradual adoption, featuring compile-time static analysis. Core capabilities include: type inference (automatic type deduction), generics (type parameterization \texttt{<T>}), union/intersection types (\texttt{string | number}, \texttt{A \& B}), and type narrowing via \\ \texttt{typeof}/type guards. Advanced type algebra supports mapped types \\ (\texttt{\{[K in T]: boolean\}}), conditional types (\texttt{T extends U ? X : Y}), and template literal types for API contracts;
    \item \textbf{Object-oriented features}: full support for classes, interfaces, and modular code organization;
    \item \textbf{Development Tools}: rich IDE integration enabling precise code navigation, intelligent refactoring, and immediate error feedback.
\end{itemize}

Since TypeScript compiles to standard JavaScript, it runs anywhere JavaScript does. This makes it particularly valuable for complex web applications where maintaining code quality at scale is essential.

\subsection{Svelte}
\textbf{Svelte} is a modern, \textit{reactive} JavaScript framework designed to build user interfaces with a focus on performance and simplicity \cite{svelteOverview}. It is a \textbf{component-based framework} that allows developers to create reusable UI components. Each component encapsulates its own logic, styles, and markup.

Unlike traditional frameworks that operate in the browser, Svelte shifts the work to compile time, generating highly efficient JavaScript code that directly manipulates the \textit{DOM} (Document Object Model). This approach eliminates the need for a \textit{virtual DOM}, resulting in faster load times and improved runtime performance.

Since this work makes use of \textbf{Svelte 5}, the following overview will focus on the features of this version.

\subsubsection{Component Architecture}
Components are \textbf{self-contained units of functionality}, each with its own logic, styles, and markup. This encapsulation simplifies the development process and enhances the maintainability of large-scale applications.

\subsubsection{Runes for State Management}
The way to implement reactivity in Svelte 5 is through the use of \textbf{Runes}. Simply put, Runes are keywords part of the Svelte syntax that allow the developer to define and manage reactive state variables \cite{vercelSvelte5}.

At the heart of Svelte 5's reactivity system are the \texttt{\$state}, \texttt{\$derived}, and \texttt{\$effect} runes:

\begin{itemize}
    \item \texttt{\$state}: declares a reactive variable. Whenever a \texttt{\$state} variable's value changes, Svelte \textit{reactively} updates the value anywhere it is used in a component;
    \item \texttt{\$derived}: creates reactive values based on other \texttt{\$state} or \texttt{\$derived} values. When a dependency changes, Svelte marks the \texttt{\$derived} value for recalculation upon its next read;
    \item \texttt{\$effect}: runs code in response to state changes (e.g., rendering to a \texttt{<canvas>} or interacting with an external library).
\end{itemize}

There are also other runes available (e.g. \texttt{\$props}, \texttt{\$bindable}, \texttt{\$inspect}), but these are not essential and thus will not be discussed here for brevity.

\subsubsection{Template Syntax}
Svelte's template syntax is a declarative language that allows developers to define the structure and behavior of their components. It is similar to HTML, but with additional features that enable reactivity and data binding. Some examples are:
\begin{itemize}
    \item \texttt{\{\#if ...\}}: conditionally renders content based on a boolean expression;
    \item \texttt{\{\#each ...\}}: iterates over an iterable and renders content for each item.
\end{itemize}

\subsubsection{SvelteKit}
\textbf{SvelteKit} is a framework built on top of Svelte that helps developers build complete web applications. While Svelte is focused on creating user interface components, SvelteKit provides the additional infrastructure needed for full application development.

At its core, SvelteKit offers:
\begin{itemize}
    \item \textbf{Routing}: built-in router that updates the UI when links are clicked;
    \item \textbf{Performance Optimizations}: loads only minimal required code and offers build optimizations;
    \item \textbf{Flexible Rendering}: supports server-side rendering (SSR), client-side rendering, and prerendering at build time;
    \item \textbf{Development Experience}: leverages Vite \cite{viteOfficial} - a JavaScript build tool providing a fast and efficient development server - for fast development with Hot Module Replacement (HMR);
    \item \textbf{Advanced Features}: includes offline support, page preloading, and image optimization.
\end{itemize}

One easy way to think of SvelteKit is as Svelte's companion for building complete applications, much like Next.js is to React or Nuxt is to Vue. While Svelte excels at creating interactive UI components, SvelteKit handles everything else you need to ship a modern web application.

\subsection{Containerization in Modern Development}
\textbf{Containerization} is a lightweight virtualization technology that packages applications with their dependencies into standardized, isolated units called containers \cite{circleciContainerizationBenefits}. Unlike traditional virtual machines, containers share the host operating system kernel while maintaining process isolation, offering several key benefits:

\begin{itemize}
    \item \textbf{Portability:} "write once, run anywhere" capability enables consistent behavior across environments from development to production;
    \item \textbf{Resource Efficiency:} shared OS kernel and elimination of virtual hardware overhead enables 3-4x greater container density per host compared to VMs;
    \item \textbf{DevOps Agility:} rapid container spin-up (often $<$1s) supports continuous integration/delivery pipelines;
    \item \textbf{Security Through Isolation:} kernel-level namespaces/cgroups prevent \\ container-to-container interference while limiting host system exposure;
    \item \textbf{Microservice Readiness:} natural fit for decomposing monolithic applications into independently scalable components.
\end{itemize}

Modern tools like Docker and Kubernetes implement these patterns through declarative configurations and orchestration systems, enabling deployment strategies discussed in Chapter 4.

    \newpage

\section{User Interface Design Principles}

\textbf{User Interface (UI) Design Principles} \cite{wikipediaUIPrinciples} are fundamental guidelines that help designers create intuitive, efficient, and user-friendly digital interfaces. These principles are rooted in cognitive psychology, human-computer interaction research, and years of design best practices. \cite{nielsen1994usability, nielsen2000designing, norman2014design}

In the following, we will discuss some crucial UI Design Principles that have been pivotal in the development of this work.

\subsection{Simplicity}

Simplicity in UI design is often encapsulated by the famous quote from Antoine de Saint-Exupéry \cite{goodreadsPerfectionQuote}: 

\begin{quote}
    \textit{"Perfection is achieved, not when there is nothing more to add, but when there is nothing left to take away."}
\end{quote}

This principle advocates for designs that are clear, straightforward, and free from unnecessary complexity. It focuses on creating interfaces that users can understand and navigate intuitively without requiring extensive learning or instruction.

At its core, \textbf{simplicity} in UI design aims to reduce cognitive load \cite{mediumUIPrinciples}. Every element, interaction, and piece of information presented to the user requires mental processing. By simplifying the interface, we reduce the mental effort required to use it, making the experience more enjoyable and efficient.

Implementing simplicity is challenging, but there are at least some key aspects to consider. For example, being \textbf{minimalistic} (e.g., limited number of fonts, colors, icons, etc.), being clear about the \textbf{information architecture} (e.g., grouping related items together logically, using white spaces effectively), \textbf{streamlining the functionality} (e.g. reducing the number of actions the user can perform, breaking complex tasks into smaller steps, etc.), and using \textbf{clear communication}.

\subsection{Responsiveness}
Responsiveness has become a cornerstone of modern UI design, particularly with the proliferation of diverse screen sizes and resolutions. This principle ensures interfaces \textit{adapt} seamlessly to various devices and screen orientations.

Designers achieve this through fluid layouts that adjust to available screen widths, relative units combined with CSS media queries to fine-tune typography and spacing, and flexible media elements that scale appropriately. Common implementations use responsive web design frameworks like Bootstrap and Tailwind.

\newpage

\subsection{Progressive Disclosure}
\textbf{Progressive disclosure} involves revealing interface features gradually, initially showing only essential options while hiding advanced or rarely used functionality.

This approach creates less intimidating interfaces that are easier to navigate, particularly for novice users (a primary target of this work). It allows users to focus on the task at hand without being overwhelmed by a multitude of options. As users gain familiarity or need advanced features, additional functionality is progressively revealed. Examples of this principle can be seen in software like Adobe Photoshop, where advanced tools are hidden under menus or toolbars.

\subsection{Consistency}
\textbf{Consistency} involves using once-adopted solutions in an \textit{unmodified} form across different interface contexts. This principle reduces cognitive load by making interfaces \textit{predictable}, which accelerates learning and enhances user comfort.

A consistent interface primarily affects three aspects:
\begin{itemize}
    \item Terminology (language consistency);
    \item Visual design (aesthetic consistency);
    \item Interaction patterns (functional consistency).
\end{itemize}

To achieve consistency, designers should use standardized components and elements, such as buttons, forms, and icons (e.g. Skeleton UI toolkit for Svelte). A consistent color palette and typography are also crucial in creating a cohesive design language. Guidelines should be established for color usage, including color contrast ratios for accessibility, as well as for typography, including font families, sizes, and line spacing.

\subsection{Accessibility}
Accessibility ensures software can be used by people with diverse abilities, including those with visual, auditory, motor, or cognitive impairments. It focuses on creating inclusive designs for all users.

Key accessibility considerations include:
\begin{itemize}
    \item \textbf{Sufficient color contrast}: be sure that text and important elements meet WCAG guidelines \cite{w3cWCAG};
    \item \textbf{Keyboard navigation}: ensure that all functionality is available via keyboard navigation;
    \item \textbf{ARIA Implementation}: implement ARIA attributes to convey meaning and functionality to screen readers without altering the visual UI;
    \item \textbf{Form labels}: design forms with clear labels and error messages;
    \item \textbf{Captions and transcripts}: provide captions and transcripts for audio and video content.
\end{itemize}
    
    % System Design
    \chapter{VoxLogicA UI}

We can now discuss the main contribution of this thesis: the \textit{VoxLogicA UI} project.

We developed VoxLogicA UI using modern web technologies while adhering to established user interface design principles. We aimed to create an interface that would be intuitive, efficient, and accessible for both medical professionals and researchers in neuroimaging, while providing full access to VoxLogicA's powerful analytical capabilities.

In the following sections, we will discuss the motivations behind the need for a modern UI for VoxLogicA, and then present the architecture of the project.

\section{The Need for a Modern UI for VoxLogicA}

In the realm of neuroimaging, a variety of tools have been developed to assist researchers and clinicians in visualizing and analyzing brain data \cite{man2015bioinformaticsToolsNeuroimaging}. Popular software like \href{https://www.osirix-viewer.com/}{OsiriX} (2004), \href{https://www.slicer.org/}{3D Slicer} (1998), \href{https://afni.nimh.nih.gov/}{AFNI} (1994), MRICron (2006), and \href{https://www.brainvoyager.com/}{BrainVoyager} (1997) generally supports various file formats and provides tools for visualization, manual editing, and automatic segmentation. However, despite their widespread use and continuous updates, many of these tools still rely on outdated interfaces designed with paradigms from their original releases.

One notable exception in the field is \href{https://nordicneurolab.com/}{nordicneurolab} and their \href{https://www.nordicmediva.com/}{nordicmediva} software, which has gained recognition for its user-friendly interface and comprehensive range of products and services, including hardware and software, for the acquisition and analysis of fMRI data. However, with licenses costing several thousand euros per year, this software is not easily accessible, particularly for smaller research institutions and individual practitioners.

\begin{figure}[h]
    \centering
    \includegraphics[width=0.5\textwidth]{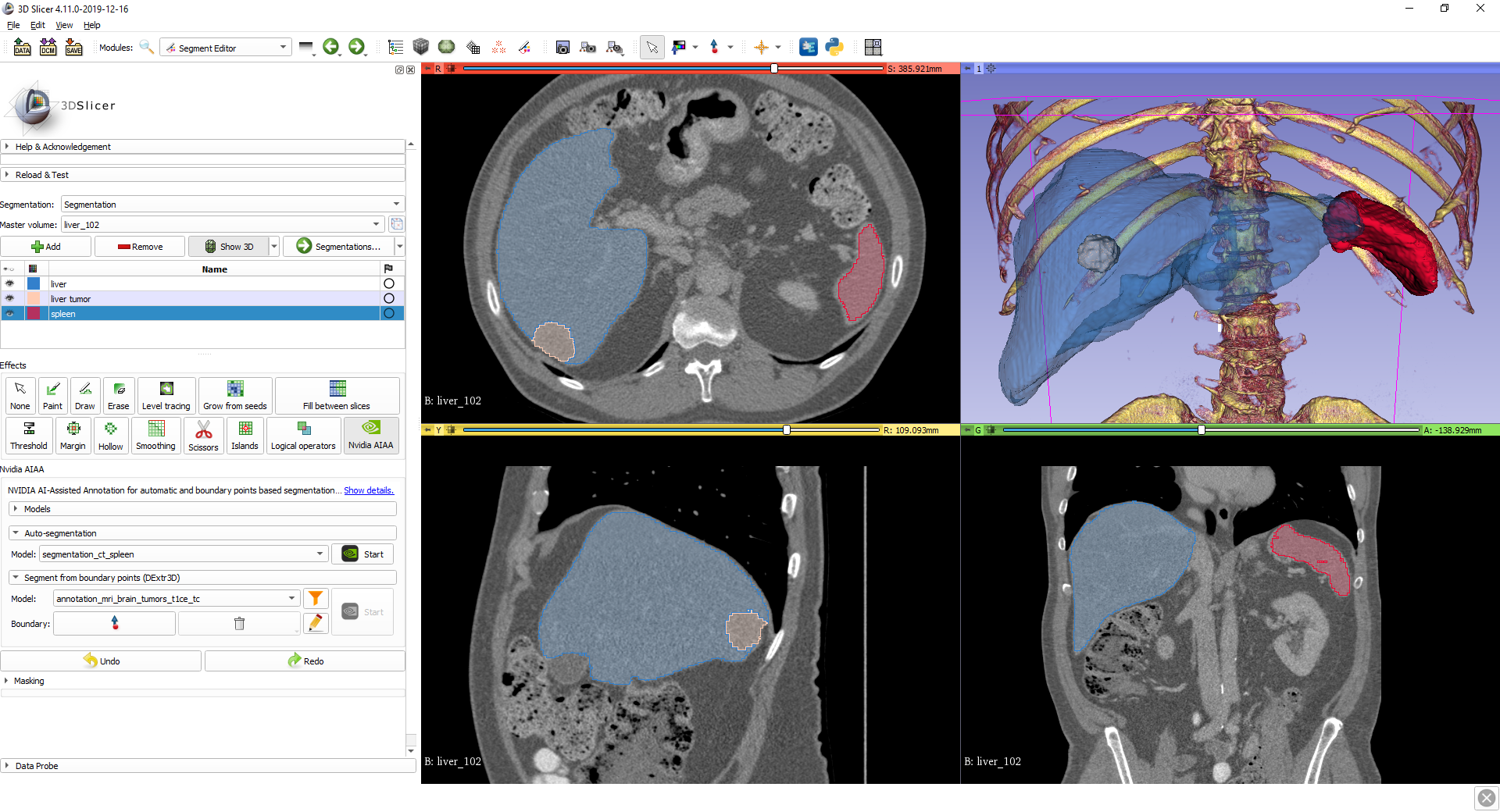}
    \caption{Screenshot of 3D Slicer.}
    \label{fig:3dslicer}
\end{figure}

As discussed in the background knowledge, the tool VoxLogicA was published in 2019 as a Spatial Model Checker specific for medical image analysis. A prototypical user interface, called VoxLogicA Explorer, was released in 2022 \cite{guiVoxlogica2022} for testing and research purposes, in order to investigate the usability and cognitive load of logic-based methods in imaging.

This first iteration represented an important step in the field of Spatial Model Checking, providing essential functionality including:

\begin{itemize}
    \item Basic image visualization capabilities;
    \item A code editor with syntax highlighting for writing VoxLogicA specifications;
    \item The ability to execute analyses directly from the interface;
    \item Visualization of analysis results overlaid on the original images.
\end{itemize}

This initial iteration of the UI, while functional, lacked modern design elements and usability features that are now standard in web applications.

User tests performed with the first UI collected valuable suggestions for improvement. For example, the relationship between the layers displayed in the layers column and those computed by the VoxLogicA procedure in the code column was not entirely clear to all users. Additionally, users experienced difficulties when adding new layers and comparing them. The automatic color selection was also criticized, with users requesting the ability to manually choose their preferred colors.

Finally, being a first experimental prototype, the UI was developed in pure HTML, CSS, and JavaScript, which made maintenance and extension challenging. The VoxLogicA UI proposed in this work addresses these issues by incorporating user interface design principles that prioritize ease of use, accessibility, and responsiveness, developed using modern web development best practices.

\begin{figure}[ht]
    \centering
    \includegraphics[width=0.8\textwidth]{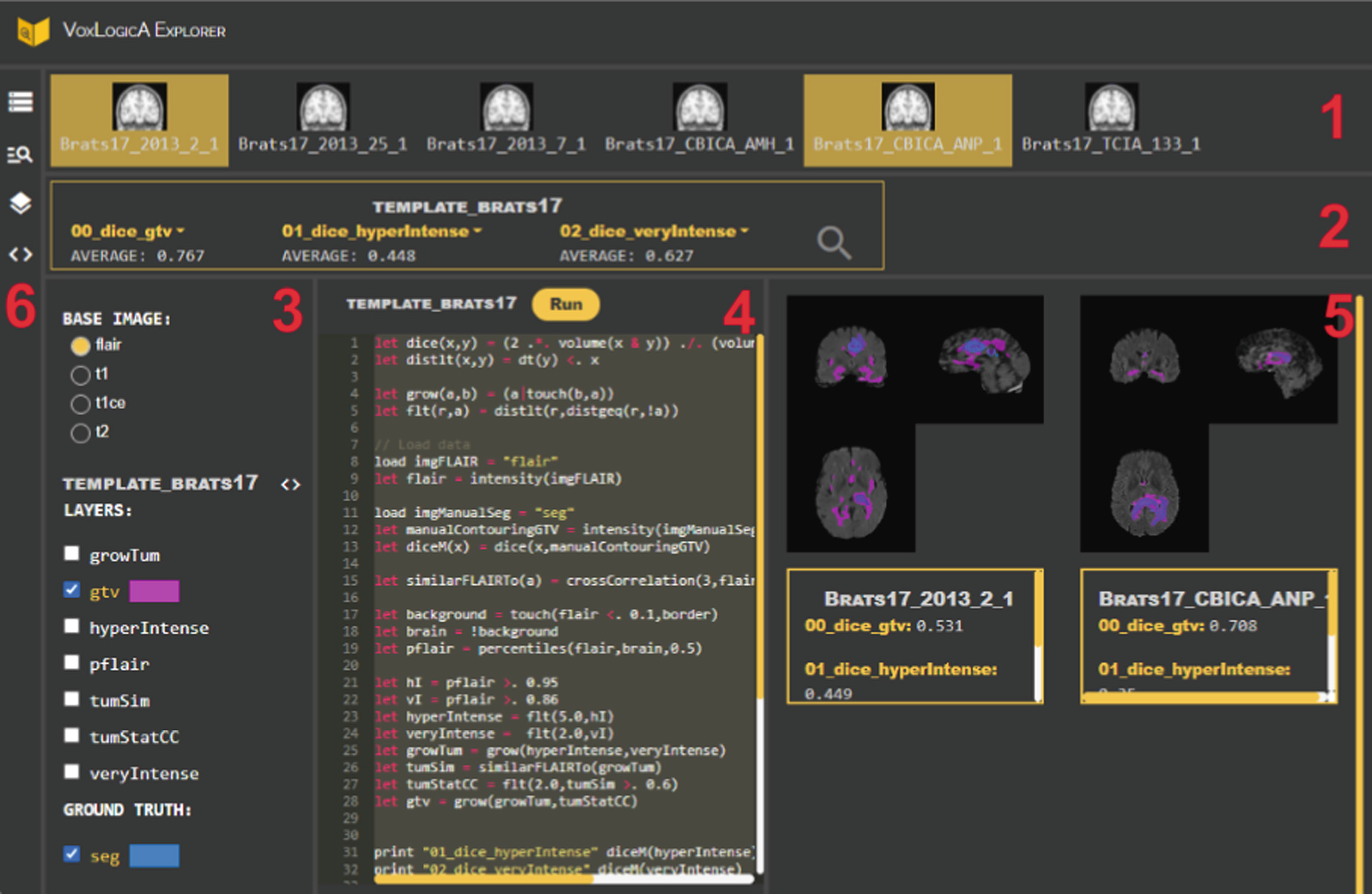}
    \caption{Screenshot of VoxLogicA Explorer interface.}
    \label{fig:voxlogica_explorer}
\end{figure}
    \section{Design Decisions}

We decided to build VoxLogicA UI as a \textbf{web application} rather than as a traditional desktop GUI. This decision was driven by two main practical considerations:
\begin{itemize}
    \item \textbf{Real-world constraints in healthcare environments:} hospitals and medical institutions often have strict IT policies that make installing external software a lengthy, bureaucratic process - sometimes taking months for approval. A web-based solution elegantly sidesteps this problem: users only need a modern web browser, which is already approved and installed in most settings;
    \item \textbf{Operating system independence:} while Windows dominates clinical settings, research labs often use a mix of macOS and Linux systems. A web architecture eliminates the need to maintain multiple platform-specific builds and installers, ensuring researchers can access the tool regardless of their preferred OS.
\end{itemize}

For our main framework, we selected \textbf{Svelte} after evaluating several alternatives including React and Vue. Svelte's compiler-based approach results in smaller bundle sizes (typically 3-4x smaller than React) and better runtime performance due to the absence of a virtual DOM. We paired it with \textbf{Skeleton UI}, which offers comprehensive component libraries and consistent theming capabilities.

For the visualization of 3D medical images, we chose \textbf{NiiVue}, which provides a performant WebGL-based rendering engine capable of handling large volumetric datasets, with native support for a wide range of medical imaging formats (e.g., NIfTI) without requiring any server-side conversion.

% \vinc{Bisogna comunque essere chiari nel dire che il paradigma MVVM è comunque mediato anche dal paradigma a componenti di svelte, e che quindi in un certo senso c'è una ibridazione fra le due cose (non tutte le variabili di stato sono sempre nel model?). Personalmente lo considererei un fatto di nessuna rilevanza ma una footnote ci potrebbe anche stare.}

For our architectural pattern, we implemented the \textbf{Model-View-ViewModel (MVVM)} pattern\footnote{In practice, Svelte's component model allows localized state management within components, creating a hybrid approach where some transient UI state may reside outside the formal Model layer while maintaining core business logic separation.}. This choice was driven by several factors: (1) the pattern's natural fit with Svelte's reactive paradigm, (2) the need to maintain clear separation between the complex image processing logic and the UI layer, and (3) the improved testability it offers through isolated layers.

The entire application is containerized using \textbf{Docker}, ensuring consistent deployment across different environments and simplifying the installation process for both server and local deployment.
    \newpage
\section{Project Architecture}

Before diving into the architecture, it is essential to understand five key domain concepts:

\begin{itemize}
    \item \textbf{Datasets:} collections of related medical imaging data, typically grouped by study or condition (e.g. BraTS, ADNI, etc.);

    \item \textbf{Cases:} individual subject/patient scans within a dataset. Each case is a folder containing a complete set of imaging data for one subject. For example, in the BraTS dataset:
    
    \begin{minipage}{\textwidth}
        \dirtree{%
        .1 Dataset: BraTS2019.
        .2 Case: patient\_001.
        .2 Case: patient\_002.
        .2 Case: patient\_003.
        }
    \end{minipage}

    \item \textbf{Layers:} a specific medical image volume associated with a case. In the UI, these come in two main categories:
    \begin{itemize}
        \item \textit{Source Layers:} these are the original layers present in dataset cases (e.g. T1, T2, FLAIR, and manual segmentations);
        \item \textit{Derived Layers:} these are new layers created with VoxLogicA in the UI (e.g. automatic tumor segmentation and intermediate results from spatial model checking).
    \end{itemize}

    The source layers can be found in the dataset cases, while the derived layers are stored in the workspace directory (see below). For example, the source layers for patient\_001 could be:

    \begin{minipage}{\textwidth}
        \dirtree{%
        .1 Case: patient\_001.
        .2 Layer: t1.nii.gz (T1-weighted MRI scan).
        .2 Layer: t2.nii.gz (T2-weighted MRI scan).
        .2 Layer: seg.nii.gz (Manual segmentation).
        }
    \end{minipage}

    \item \textbf{Workspaces:} a complete saved state of a user's analysis session, including both the UI state and analysis outputs. It maintains all data necessary to restore the working environment and access analysis results.

    The workspace directory stored in the server contains:
    \begin{itemize}
        \item \textbf{Session State:} which datasets and cases are open, how every loaded layer is visualized (z-index, color maps, opacity...), UI configuration (dark mode, sidebar states...), and the current analysis script;
        \item \textbf{Analysis Results:} generated layers from ImgQL (Image Query Language) scripts, execution logs and outputs, and metadata.
    \end{itemize}

    \item \textbf{Example scripts:} programs written in ImgQL that define spatial model checking operations. These are provided as examples in the UI, and can be used as starting points for new analyses.
\end{itemize}

\newpage

The architecture follows a layered pattern, with each layer having specific responsibilities and communicating only with adjacent layers.

\begin{figure}[ht]
    \centering
    \begin{tikzpicture}[
        % Node styles
        block/.style={rectangle, draw, text width=2.8cm, minimum height=0.8cm, 
                      align=center, font=\footnotesize},
        layer/.style={rectangle, draw, dashed, inner sep=8pt},
        arrow/.style={->, >=latex},
        note/.style={font=\scriptsize, align=center, text width=3.2cm},
        node distance=0.3cm
    ]

    % Frontend Layer
    \node[layer, fill=blue!5] (frontend_layer) {
        \begin{tikzpicture}[node distance=0.2cm]
            % Main functional areas
            \node[block] (navigation) {Navigation \& Management};
            \node[block, right=1.2cm of navigation] (visualization) {Visualization\\\hspace{1cm}};
            \node[block, right=1.2cm of visualization] (analysis) {Spatial Model Checking};
            
            % Description notes
            \node[note, below=0.1cm of navigation] {Dataset browser\\Case management\\Workspace handling};
            \node[note, below=0.1cm of visualization] {Medical image viewers\\Layer management};
            \node[note, below=0.1cm of analysis] {Script editor (ImgQL)\\VoxLogicA runner};
        \end{tikzpicture}
    };

    % State Management Layer
    \node[layer, below=0.8cm of frontend_layer, fill=green!5] (state_layer) {
        \begin{tikzpicture}[node distance=0.2cm]
            % ViewModels
            \node[block] (session_vm) {Session VM};
            \node[block, right=1.2cm of session_vm] (dataset_vm) {Dataset VM};
            \node[block, right=1.2cm of dataset_vm] (case_vm) {Case VM};
            \node[block, below=0.2cm of session_vm] (layer_vm) {Layer VM};
            \node[block, below=0.2cm of dataset_vm] (run_vm) {Run VM};
            \node[block, below=0.2cm of case_vm] (ui_vm) {UI VM};
        \end{tikzpicture}
    };

    % Repository Layer
    \node[layer, below=0.8cm of state_layer, fill=orange!5] (repository_layer) {
        \begin{tikzpicture}[node distance=0.2cm]
            \node[block] (dataset_repo) {Dataset Repository};
            \node[block, right=1.2cm of dataset_repo] (workspace_repo) {Workspace Repository};
            \node[block, right=1.2cm of workspace_repo] (script_repo) {Script Repository};
        \end{tikzpicture}
    };

    % Backend Interface Layer
    \node[layer, below=0.8cm of repository_layer, fill=yellow!5] (backend_layer) {
        \begin{tikzpicture}[node distance=0.2cm]
            % Server Routes
            \node[block] (dataset_api) {Dataset Routes};
            \node[block, right=1.2cm of dataset_api] (workspace_api) {Workspace Routes};
            \node[block, right=1.2cm of workspace_api] (script_api) {Script Routes};
            \node[block, below=0.2cm of workspace_api] (run_api) {Run Routes};
        \end{tikzpicture}
    };

    % Storage Layer
    \node[layer, below=0.8cm of backend_layer, fill=red!5] (storage_layer) {
        \begin{tikzpicture}[node distance=0.2cm]
            \node[block] (datasets) {Datasets};
            \node[block, right=1.2cm of datasets] (scripts) {Scripts};
            \node[block, right=1.2cm of scripts] (workspaces) {Workspaces};
        \end{tikzpicture}
    };

    % Add arrows between layers
    \draw[<->, >=latex] (frontend_layer) -- node[right] {\scriptsize State Updates} (state_layer);
    \draw[<->, >=latex] (state_layer) -- node[right] {\scriptsize Repository Calls} (repository_layer);
    \draw[<->, >=latex] (repository_layer) -- node[right] {\scriptsize API Calls} (backend_layer);
    \draw[<->, >=latex] (backend_layer) -- node[right] {\scriptsize File I/O} (storage_layer);

    \end{tikzpicture}

    \caption{VoxLogicA UI architecture overview.}
    \label{fig:voxlogica_ui_architecture}
\end{figure}
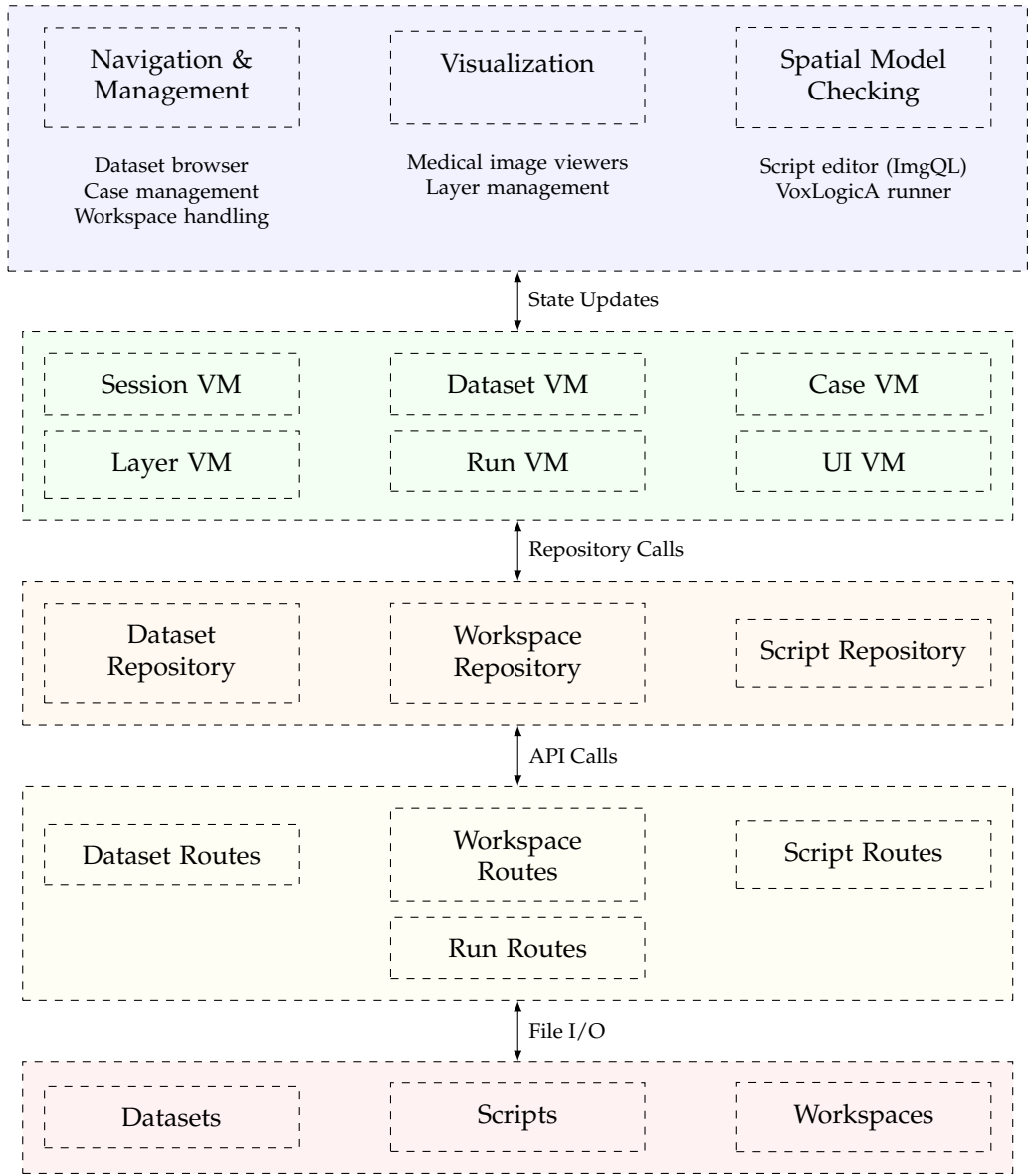

Let us examine each layer in detail.

\newpage

\subsection{View Layer}

The View Layer implements the UI components displayed in the browser, collected in three main functional areas:

\begin{itemize}
    \item \textbf{Navigation \& Management:}
    \begin{itemize}
        \item A hierarchical browser for navigating datasets, their cases, and the runs associated with each case;
        \item Search and filtering among cases and derived layers' outputs;
        \item Workspace management tools for saving and loading analysis sessions.
    \end{itemize}

    \item \textbf{Visualization:}
    \begin{itemize}
        \item Multiple synchronized NiiVue viewers, each showing different orthogonal planes (axial, sagittal, coronal) and a 3D visualization. One viewer per opened case;
        \item A synoptic panel for controlling the visibility, opacity, and color mapping of all layers, including the derived ones.
    \end{itemize}

    \item \textbf{Spatial Model Checking:}
    \begin{itemize}
        \item A fully fledged ImgQL script editor, featuring file opening/downloading, syntax highlighting and auto-completion;
        \item Controls for starting VoxLogicA analyses.
    \end{itemize}
\end{itemize}

\subsection{ViewModel Layer}

The ViewModel Layer implements the main business logic of the UI through six ViewModels:

\begin{itemize}
    \item \textbf{Session VM:} handles the workspace states and their persistence;

    \item \textbf{Dataset VM:} handles datasets loading/unloading;

    \item \textbf{Case VM:} handles cases loading/unloading;

    \item \textbf{Layer VM:} controls layers loading/unloading and their visualization settings, including visibility, colors, and z-index;

    \item \textbf{Run VM:} manages VoxLogicA analysis execution and results;

    \item \textbf{UI VM:} handles UI configuration (theme, layout, modals).
\end{itemize}

\subsection{Model Layer}

The Model Layer manages data operations via three main repositories:

\begin{itemize}
    \item \textbf{Dataset Repository:} handles dataset discovery and case/layers retrieval;

    \item \textbf{Workspace Repository:} handles workspace lifecycle, state serialization, and analysis results;

    \item \textbf{Script Repository:} handles example scripts discovery and fetching.
\end{itemize}

\subsection{Back end Layer}

The Back end Layer exposes a REST\footnote{REST (Representational State Transfer): API architecture using HTTP methods (GET/POST/etc.) for stateless resource operations.} API organized around the core domain concepts rather than generic CRUD\footnote{CRUD (Create, Read, Update, Delete): Basic data operations typically associated with database systems.} operations:

\begin{itemize}
    \item \textbf{Dataset Routes:} access and manage medical imaging data, following the natural hierarchy of the domain (dataset/case/layer);
    \item \textbf{Workspace Routes:} handles the workspace lifecycle;
    \item \textbf{Script Routes:} handles the fetching of the example ImgQL scripts;
    \item \textbf{Run Routes:} handles VoxLogicA execution requests and result retrieval.
\end{itemize}

\subsection{Storage Layer}

The Storage Layer represents the physical organization of data on the server's filesystem:

\begin{itemize}
    \item \textbf{Datasets:} contains the medical imaging data organized by dataset and case;
    \item \textbf{Scripts:} stores example ImgQL scripts and templates;
    \item \textbf{Workspaces:} houses workspace data including analysis results and session states.
\end{itemize}

\section{Data Flow}

To illustrate how these layers interact, let us consider a typical user workflow: loading and visualizing a medical image.

\begin{enumerate}
    \item The user selects a layer in the front end's navigation component;
    \item The Layer ViewModel requests data through the Dataset Repository;
    \item The Dataset Repository makes an API call through the Back end Interface;
    \item The Back end retrieves the NIfTI file from the Storage Layer;
    \item The data flows back up through the layers, ultimately being displayed in the visualization components.
\end{enumerate}

\begin{figure}[h]
    \centering
    \begin{tikzpicture}[
        node distance=1.2cm,
        box/.style={
            rectangle, 
            draw, 
            minimum width=1.6cm, 
            minimum height=0.7cm, 
            align=center,
            font=\footnotesize
        },
        arrow/.style={->, >=latex}
    ]
        % Nodes
        \node[box] (ui) {Front end\\UI};
        \node[box, right=of ui] (vm) {Layer\\ViewModel};
        \node[box, right=of vm] (repo) {Dataset\\Repository};
        \node[box, right=of repo] (api) {Back end\\API};
        \node[box, right=of api] (storage) {Storage\\Layer};
        
        % Forward arrows
        \draw[arrow] (ui) -- node[above, font=\scriptsize] {1. select} (vm);
        \draw[arrow] (vm) -- node[above, font=\scriptsize] {2. request} (repo);
        \draw[arrow] (repo) -- node[above, font=\scriptsize] {3. call} (api);
        \draw[arrow] (api) -- node[above, font=\scriptsize] {4. fetch} (storage);
        
        % Return arrows
        \draw[arrow] ([yshift=-0.3cm]storage.west) -- node[below, font=\scriptsize] {data} ([yshift=-0.3cm]api.east);
        \draw[arrow] ([yshift=-0.3cm]api.west) -- node[below, font=\scriptsize] {data} ([yshift=-0.3cm]repo.east);
        \draw[arrow] ([yshift=-0.3cm]repo.west) -- node[below, font=\scriptsize] {data} ([yshift=-0.3cm]vm.east);
        \draw[arrow] ([yshift=-0.3cm]vm.west) -- node[below, font=\scriptsize] {data} ([yshift=-0.3cm]ui.east);
    \end{tikzpicture}
    \caption{Data flow when loading a layer.}
    \label{fig:data_flow}
\end{figure}
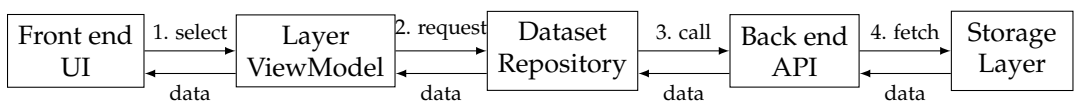

We believe that this layered architecture provides a clear separation of concerns, making the code easier to maintain and test, while also being easy to scale.

In the next chapter, we will delve into the details of the project's implementation.
    
    % % Implementation
    \chapter{Implementation}

Having established the architectural foundations and design decisions in Chapter 3, we can now delve into the concrete implementation of VoxLogicA UI. This chapter will detail how the theoretical architecture was translated into code, exploring the key implementation choices and technical solutions adopted.

\section{Codebase Overview}

The project is built using TypeScript to ensure type safety throughout the application, and leverages SvelteKit's file-based routing system for structural organization.

The repository follows a clear organizational pattern, with the main source code residing in the \texttt{src/} directory. The codebase is organized into several key directories that closely mirror the layered architecture discussed in Chapter 3:

\begin{itemize}
    \item \texttt{src/lib/}: contains the core application logic, with the following subdirectories:
    \begin{itemize}
        \item \texttt{components/}: contains UI components organized by feature area (common, layers, navigation, run, viewers);
        \item \texttt{viewmodels/}: contains ViewModels, the bridge between the Model and View layers;
        \item \texttt{models/}: contains type definitions and Models.
    \end{itemize}
    
    \item \texttt{src/routes/}: contains the implementation of the \textit{Back end} layer through SvelteKit's server routes, thus organized by API endpoints (datasets, \\ workspaces, run, scripts);
    
    \item \texttt{src/tests/}: contains comprehensive test suites for models and viewmodels.
\end{itemize}

This organization aims to follow current best practices for maintainability and scalability \cite{svelteKitDesignPattern}, maintaining a clear separation between front end and back end while keeping them in a single codebase (a pattern enabled by SvelteKit's unified application structure \cite{svelteKitProjectStructure}).

    \section{Development Environment}

To ensure consistency across different development environments and simplify the onboarding of new contributors, we utilized \textbf{Docker containers} for development. This approach provides several benefits:

\begin{itemize}
    \item \textbf{Environment reproducibility:} the development environment is fully specified in code, eliminating "works on my machine" issues;
    \item \textbf{Automated setup:} new developers can start working on the project with a single command, as the container automatically:
    \begin{itemize}
        \item Sets up the correct Node.js environment;
        \item Installs all dependencies;
        \item Downloads and configures the VoxLogicA binary;
        \item Sets up recommended VS Code extensions for development.
    \end{itemize}
    \item \textbf{Parity with production:} the development environment closely mirrors the production environment, reducing deployment-related issues;
    \item \textbf{Enhanced security:} the application runs in an isolated container environment as a non-root user (least privilege), and the resources available to the container are limited.
\end{itemize}

Developers are encouraged to use \textbf{VS Code} with the \textbf{Dev Containers} extension, as the project includes a \texttt{devcontainer.json} file that automatically configures the development environment when opening the project in VS Code.

The Docker build processes we defined are specified through a \textbf{multi-stage Docker build process} \cite{dockerMultiStage}, to optimize both development and production environments:

\begin{itemize}
    \item \textbf{Builder stage:} handles application compilation and building;
    \item \textbf{Production stage:} creates a minimal runtime environment with only essential components.
\end{itemize}

This approach has the benefit of generating smaller images, while maintaining full development capabilities.

\subsection{Code Quality Assurance}
% \vinc{le sezioni 4.2.1 e 4.3 appaiono alla vista come una lista di file e API, forse potrebbero stare in fondo al capitolo o come appendice? (naturalmente estraendo l'info "non tabellare" e lasciandola dove è)} % ANTONIO: per il momento non saprei come ri-organizzare il tutto, direi di valutare in base al tempo rimasto.
Several tools and practices are employed to maintain high code quality:

\begin{itemize}
    \item \textbf{Static Type Checking:} TypeScript provides compile-time type checking, helping to catch errors early in development;
    
    \item \textbf{Code Formatting:} Prettier ensures consistent code style across the codebase, configured through:
    \begin{itemize}
        \item \texttt{.prettierrc} for style rules;
        \item \texttt{.prettierignore} for excluding specific files/paths.
    \end{itemize}
    
    \item \textbf{Continuous Integration:} a GitHub Actions workflow (\texttt{.github/workflows/ci.yml}) automatically:
    \begin{itemize}
        \item Runs the test suite, ensuring our models and viewmodels keep behaving as expected;
        \item Checks code formatting;
        \item Verifies type correctness;
        \item Ensures the build process succeeds.
    \end{itemize}
    
    \item \textbf{Development Scripts:} a comprehensive set of npm scripts facilitates common development tasks:
    \begin{itemize}
        \item \texttt{yarn check} for type checking;
        \item \texttt{yarn format} for code formatting;
        \item \texttt{yarn test} for running tests;
        \item \texttt{yarn test:coverage} for generating coverage reports.
    \end{itemize}
\end{itemize}
    \section{Back end}

For the back end, we focused on providing a clean REST API that serves as the interface between the front end and the underlying file system and VoxLogicA binary.

The API follows a hierarchical RESTful structure, where endpoints are organized around the core domain concepts identified in the system architecture: datasets, cases, layers, workspaces, and scripts.

In particular:

\begin{itemize}
    \item \texttt{/datasets} (GET)
        \begin{itemize}
            \item Purpose: retrieves all available datasets;
            \item Returns: array of Dataset objects;
        \end{itemize}
        
    \item \texttt{/datasets/[dataset]} (GET)
        \begin{itemize}
            \item Purpose: retrieves information about a specific dataset;
            \item Returns: Dataset object containing name;
        \end{itemize}
        
    \item \texttt{/datasets/[dataset]/cases} (GET)
        \begin{itemize}
            \item Purpose: retrieves all cases within a dataset;
            \item Returns: array of Case objects containing name and path;
        \end{itemize}
        
    \item \texttt{/datasets/[dataset]/cases/[case]/layers} (GET)
        \begin{itemize}
            \item Purpose: retrieves all layers for a specific case;
            \item Returns: array of Layer objects containing name and path;
        \end{itemize}
    
    \newpage
    
    \item \texttt{/datasets/[dataset]/cases/[case]/layers/[layer]} (GET)
        \begin{itemize}
            \item Purpose: downloads a specific layer file;
            \item Returns: binary file stream (NIFTI format);
        \end{itemize}
        
    \item \texttt{/run} (POST)
        \begin{itemize}
            \item Purpose: executes VoxLogicA script on specified cases (the integration with VoxLogicA has been made through a dedicated system call directly leveraging the VoxLogicA binary);
            \item Accepts: workspaceId, scriptContent, and array of cases;
            \item Returns: array of Run results;
        \end{itemize}
        
    \item \texttt{/scripts} (GET)
        \begin{itemize}
            \item Purpose: retrieves all available example scripts;
            \item Returns: array of ExampleScript objects;
        \end{itemize}
        
    \item \texttt{/scripts/[script]} (GET)
        \begin{itemize}
            \item Purpose: retrieves the content of a specific script;
            \item Returns: plain text content of the script;
        \end{itemize}
        
    \item \texttt{/workspaces} (POST)
        \begin{itemize}
            \item Purpose: creates a new workspace or clones an existing one;
            \item Accepts: optional sourceId (workspaceId to clone from) and workspace data (name, initial configuration, etc.);
            \item Returns: new Workspace object;
        \end{itemize}
        
    \item \texttt{/workspaces/[workspaceId]} (GET, PUT, DELETE)
        \begin{itemize}
            \item GET: retrieves workspace information;
            \item PUT: updates workspace information;
            \item DELETE: removes workspace and all associated data;
            \item Returns: workspace object (GET), success status (PUT/DELETE);
        \end{itemize}
        
    \item \texttt{/workspaces/[workspaceId]/runs} (GET)
        \begin{itemize}
            \item Purpose: retrieves all runs for a workspace;
            \item Returns: array of Run objects;
        \end{itemize}
        
    \item \texttt{/workspaces/[workspaceId]/[caseId]/[runId]/layers/[layer]} (GET)
        \begin{itemize}
            \item Purpose: downloads a specific output layer from a run;
            \item Returns: binary file stream (NIFTI format);
        \end{itemize}
\end{itemize}

All responses follow a standard structure, with success responses using appropriate HTTP status codes (200, 201) and error responses including descriptive messages along with corresponding status codes (400, 404, 500).

% TODO: maybe go a bit more in detail on the security aspect of the API, and how we implemented it

\subsection{Data Management Implementation}

The application implements a structured filesystem hierarchy to manage different types of data. While Chapter 3 introduced the conceptual organization, the actual implementation uses \textbf{environment variables} to allow flexible and secure configuration of data locations:

\begin{itemize}
    \item \texttt{DATASET\_PATH}: houses medical imaging datasets;
    \item \texttt{SCRIPTS\_PATH}: stores example ImgQL scripts;
    \item \texttt{WORKSPACES\_PATH}: contains user workspace data;
    \item \texttt{VOXLOGICA\_BINARY\_PATH}: points to the VoxLogicA executable.
\end{itemize}

This configuration system defaults to the \texttt{static} directory in development and \texttt{/data} in production (Docker), ensuring consistent behavior across environments while maintaining flexibility for different deployment scenarios.

\subsubsection{Workspace State Management}

Workspaces implement a sophisticated state persistence strategy that goes beyond simple file storage. Each workspace maintains:

\begin{itemize}
    \item \textbf{Atomic Updates}: state changes are written to temporary files first, then atomically renamed to ensure consistency even during system crashes; 
    \item \textbf{Differential Storage}: only modified layers are saved, with references to source data for unchanged layers, minimizing storage requirements while avoiding the creation of a dependency tree between workspaces;
    \item \textbf{Dependency Tracking}: each derived layer maintains metadata about its source layers and the ImgQL script that generated it.
\end{itemize}

We also implemented a workspace cloning mechanism through the \texttt{/workspaces} endpoint, enabling collaboration between colleagues. Researchers and clinicians working together can clone an existing workspace and modify it, while the original workspace remains unchanged. This matches clinical workflows where residents analyze attending physicians' cases without modifying original diagnoses.

Our solution addresses three critical challenges in medical collaboration: 
\begin{itemize}
    \item \textbf{Data Sovereignty}: in hospital environments where PACS systems often reside on air-gapped networks, our clone-and-modify approach creates derivative workspaces \textit{without allowing to export original patient data};
    \item \textbf{Audit Compliance}: the immutable workspace ID system enables traceability back to source data through a directed acyclic graph (DAG) of workspace relationships, satisfying GDPR/HIPAA requirements for provenance tracking;
    \item \textbf{Collaboration Friction}: unlike traditional DICOM sharing that requires exporting entire studies (often 100+ MB per case), our method enables peer review through \textit{workspace snapshots} containing just:
    \begin{itemize}
        \item Final segmentation masks (NIfTI $\sim$2-5MB);
        \item Quantitative measurements (JSON $<$1KB);
        \item Execution log with timestamps.
    \end{itemize}
\end{itemize}

We considered and discarded other solutions as well:
\begin{itemize}
    \item \textbf{File Sharing via Email/USB}: would require moving sensitive data off secured systems, violating hospital IT policies;
    \item \textbf{Full Dataset Duplication}: storage-prohibitive for whole-body MRI studies (typically 500MB-2GB/case).
\end{itemize}

\subsubsection{Data Loading}

We implemented \textit{lazy loading} at multiple levels to handle large datasets efficiently:

\begin{itemize}
    \item \textbf{Dataset Discovery}: only dataset and case metadata is initially loaded;
    \item \textbf{Layer Loading}: NIfTI files employ on-demand loading, prioritizing memory efficiency over access latency. While predictive prefetching based on user navigation patterns could mitigate this, we deferred implementation pending comprehensive performance profiling to balance memory utilization and responsiveness;
    % \vinc{il che però crea latenza al caricamento!; un  prefetching rallentato in quel caso potrebbe funzionare meglio}
    \item \textbf{Workspace Recovery}: when reopening a workspace, only the state file is initially loaded, with derived layers restored as needed.
\end{itemize}

\subsubsection{Error Handling}

The data management system includes error handling at multiple levels:

\begin{itemize}
    \item \textbf{Load Error Detection}: each ViewModel maintains error states to track and display loading issues;
    \item \textbf{Run Error Handling}: the \texttt{runViewModel} captures and manages VoxLogicA execution errors;
    \item \textbf{State Recovery}: the system maintains separate error states for different operations, allowing granular error reporting and recovery.
\end{itemize}   
    \newpage

\section{Front end Implementation}

The front end implementation follows a layered architecture pattern, utilizing Svelte 5's powerful features to create a robust and maintainable application. The following sections detail the implementation of the Model, ViewModel, and View layers.

\begin{figure}[h]
    \centering
    \includegraphics[width=\textwidth]{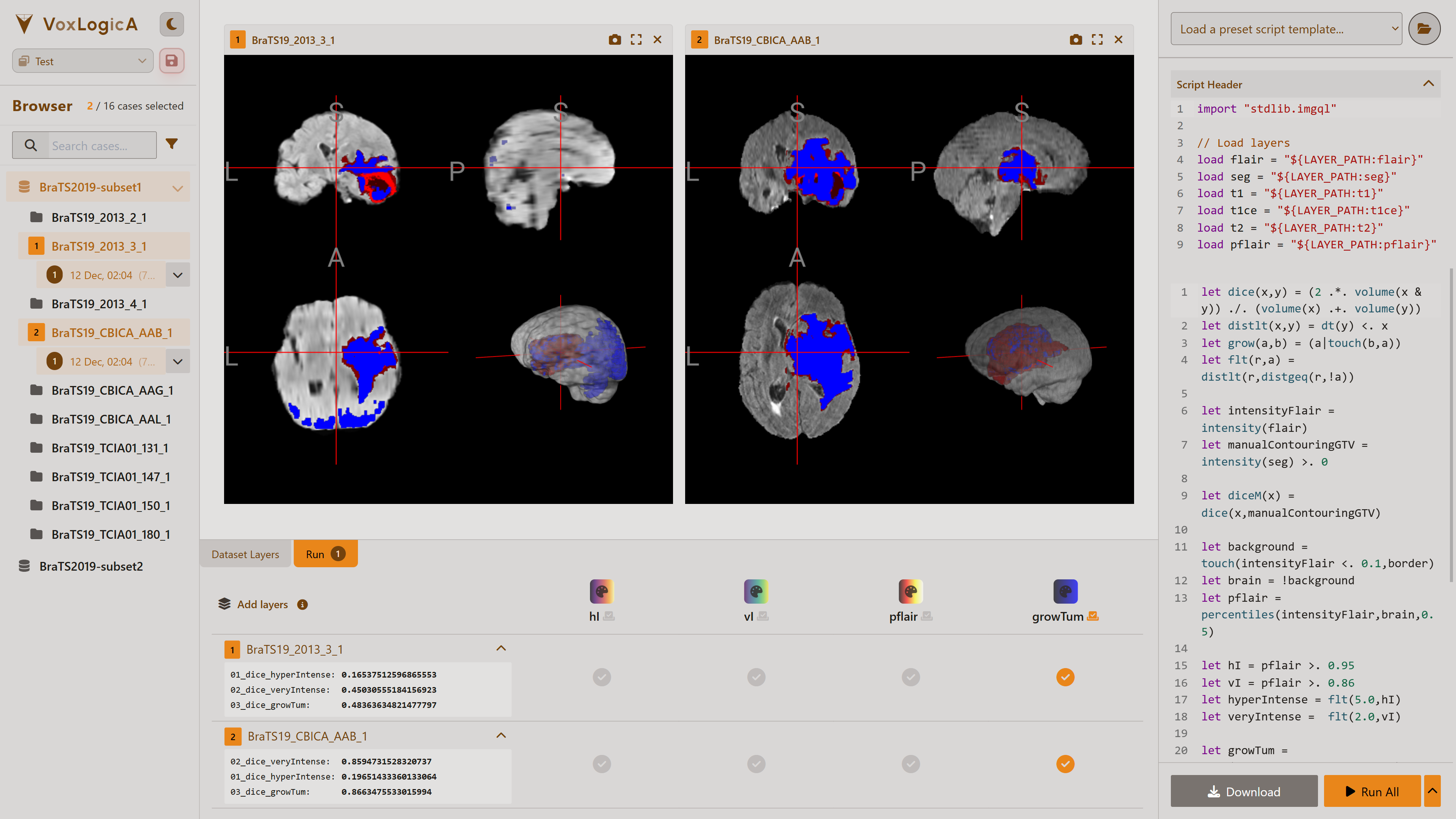}
    \caption{VoxLogicA UI interface showing the main components: browser panel (left), medical image viewers (center), layers management (bottom), and script editor (right). The interface displays two brain MRI cases with overlaid analysis results in blue.}
    \label{fig:voxlogica-ui}
\end{figure}

\subsection{Model Layer}

At the heart of VoxLogicA UI's data management lies the Model layer, which manages the application's state (including server-loaded data) through Svelte 5's runes system.

\subsubsection{Repository Pattern Implementation}

The Model layer implements a repository design pattern \cite{repositoryDesignPattern} through the \texttt{apiRepository} object, which encapsulates all API communication logic and state management.

The repository exposes operations covering:
\begin{itemize}
    \item \textbf{Dataset Operations:} \texttt{fetchDatasets()}, \texttt{fetchCases(dataset)}, and \texttt{fetchLayers(case\_)};
    \item \textbf{Workspace Operations:} CRUD operations on workspaces;
    \item \textbf{Analysis Management:} \texttt{fetchExampleScripts()} and \\ \texttt{fetchExampleScriptCode(scriptId)}.
\end{itemize}

The API invocation is implemented through a utility object that provides type-safe wrappers around the Fetch API. It offers two core functionalities: JSON data fetching with generic type parameters for compile-time type safety and text content retrieval (primarily used for ImgQL scripts).

The implementation includes comprehensive error handling through a custom \texttt{RepositoryError} class that wraps network and parsing errors while preserving the original error context. This error handling strategy ensures that API failures are properly captured and can be meaningfully handled by the application.

These elements are encapsulated in the repository module, ensuring API calls follow consistent patterns while maintaining application-wide type safety.

\subsubsection{Core State Management}

We implemented a \textbf{two-layer state management system}: global states reside in the Model layer, while ViewModels maintain component-specific states (e.g. loading states and error handling). We chose this approach since it allowed to have a single source of truth, minimizing the amount of data being passed around (and thus, improving performance), while still being flexible.

Focusing on the Model layer's state management, the Repository Pattern operates on two main global states exported by the Model layer: \texttt{loadedData} and \texttt{currentWorkspace}.

The \texttt{loadedData} state contains all server-loaded data, including:
\begin{itemize}
    \item \texttt{availableWorkspacesIdsAndNames}: list of workspaces available to the user;
    \item \texttt{datasets}: list of available medical imaging datasets;
    \item \texttt{casesByDataset}: a map of dataset names to their cases;
    \item \texttt{layersByCasePath}: a map of case paths to their layers;
    \item \texttt{runsByCasePath}: a map of case paths to their analysis runs;
    \item \texttt{exampleScripts}: list of available example ImgQL scripts.
\end{itemize}

The \texttt{currentWorkspace} state represents the active workspace configuration, containing:
\begin{itemize}
    \item Data-related states:
    \begin{itemize}
        \item \texttt{openedDatasetsNames}: currently opened datasets;
        \item \texttt{openedCasesPaths}: currently opened cases;
        \item \texttt{openedRunsIds}: currently opened analysis runs;
    \end{itemize}
    \item \texttt{lastGlobalStylesByLayerName}: saved visualization settings;
    \item Dataset layers configuration:
    \begin{itemize}
        \item \texttt{openedLayersPathsByCasePath}: which layers are open in each case;
        \item \texttt{stylesByLayerName}: visual styles for dataset layers;
    \end{itemize}
    \item \texttt{runsLayersStates}: visual styles for analysis results;
    \item UI-related states:
    \begin{itemize}
        \item \texttt{isDarkMode}: theme preference;
        \item Sidebar states (navigation, layer controls, and script editor);
        \item \texttt{fullscreenCasePath}: currently fullscreened case;
        \item \texttt{layerContext}: used to determine if the user is currently editing dataset's layers or analysis run's layers;
        \item \texttt{scriptEditor.content}: current script content.
    \end{itemize}
\end{itemize}

\subsection{ViewModel Layer}

We structured the business logic into specialized ViewModels, each handling a specific component domain. Each ViewModel follows this consistent structure:

\begin{itemize}
    \item Local state management using runes;
    % \vinc{reference to what are runes, term that should be in emphasized font the first time it is used} % ANTONIO: they are explained inside the background chapter, so I avoided re-explaining them here.
    \item Derived states computed from the global states or other derived states;
    \item Action methods that modify state;
    \item Public getters and setters.
\end{itemize}

Let us now take a look at some parts of the Case ViewModel as an example:

\begin{lstlisting}[language=JavaScript]
// Reactive states using runes
let isLoading = $state(false);
let error = $state<string | null>(null);

// Derived states (from the global states in the Model layer)
const cases = $derived.by(() => {
    return loadedData.datasets.flatMap(
        (dataset) => loadedData.casesByDataset[dataset.name] || []
    );
});

// Action methods
async function selectCase(case_: Case): Promise<void> {
	if (isSelected(case_.path)) return;
	if (!canSelectMore) {
		error = `Cannot select more than ${MAX_SELECTED_CASES} cases`;
		return;
	}
	error = null;

	try {
		await apiRepository.fetchLayers(case_);
		currentWorkspace.state.data.openedCasesPaths = [
			...currentWorkspace.state.data.openedCasesPaths,
			case_.path,
		];
	} catch (e) {
		error = e instanceof Error ? e.message : `Failed to load layers for case: ${case_.name}`;
	}
}
\end{lstlisting}

For each ViewModel, we then exported part of the states/methods through an immutable interface, and made states/variables accessible only through getters/setters.

\paragraph{Reactivity Decisions}
The ViewModel implementation considered several reactivity patterns:
\begin{itemize}
    \item \textbf{Class-based Observables:} offered familiar OOP patterns but conflicted with Svelte's compiler optimizations;
    \item \textbf{Functional Reactivity (Chosen):} leveraged Svelte's built-in \texttt{\$state} and \texttt{\$derived} runes for better compiler integration.
\end{itemize}
This choice enabled fine-grained updates while maintaining type safety through TypeScript generics.

\subsection{View Layer}

The View layer represents the user-facing components of VoxLogicA UI. We designed every interaction with the user to be intuitive and simple, guiding the user actions through tooltips and messages in every step.

The components are organized in a hierarchical structure under \texttt{src/lib/components}, with distinct verticals each using its corresponding ViewModel.

\subsubsection{Navigation Components}

We grouped the navigation components in 2 sets: browser and workspace.

The browser components provides a hierarchical view of datasets, cases, and analysis runs:
\begin{itemize}
    \item \texttt{Browser}: orchestrates the overall navigation experience;
    \item \texttt{DatasetList}: manages dataset enumeration and filtering;
    \item \texttt{CaseItem}: handles individual case representation;
    \item \texttt{RunItem}: displays analysis results;
    \item \texttt{SearchAndFilters}: provides dataset and case filtering capabilities across multiple hierarchical levels simultaneously.
\end{itemize}

The workspace component manages user sessions and analysis states:
\begin{itemize}
    \item \texttt{Workspaces}: handles workspace listing and selection;
    \item \texttt{CreateFromIdModal}: common modal for workspace creation;
    \item \texttt{WorkspacesFullScreen}: welcome screen when starting the application, prompting the user to create a workspace or load an existing one.
\end{itemize}

These components implement automatic state persistence, ensuring that user work is saved continuously.

\begin{figure}[h]
    \centering
    \begin{subfigure}[b]{0.48\textwidth}
        \centering
        \includegraphics[height=7cm]{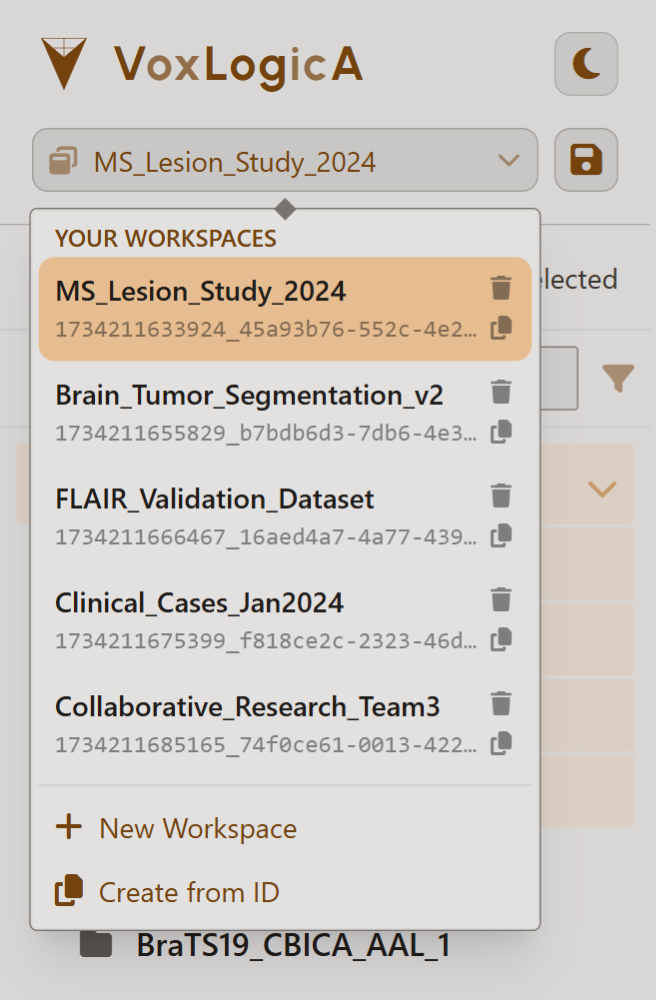}
        \caption{Workspace management interface displaying available workspaces and creation options.}
        \label{fig:workspace}
    \end{subfigure}
    \hfill
    \begin{subfigure}[b]{0.48\textwidth}
        \centering
        \includegraphics[height=7cm]{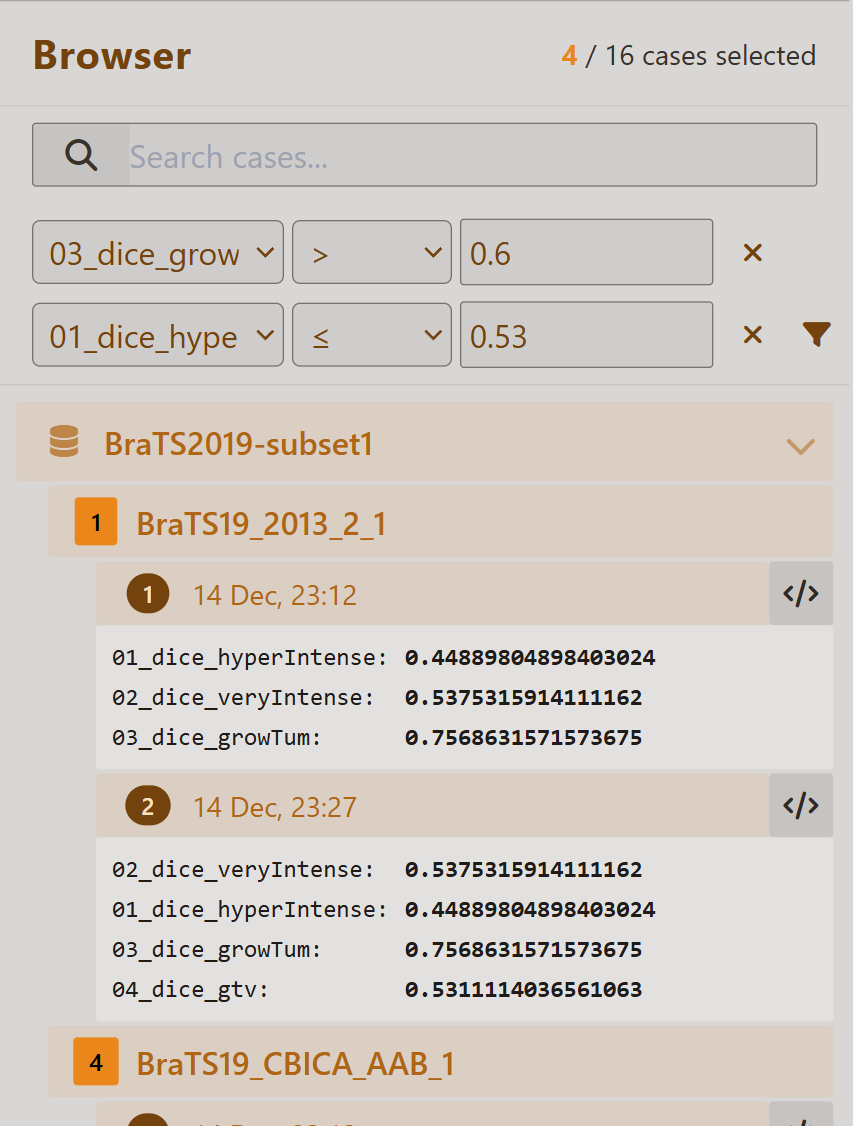}
        \caption{Browser interface showing dataset filtering and case selection with analysis results.}
        \label{fig:browser}
    \end{subfigure}
    \caption{VoxLogicA UI navigation components.}
    \label{fig:navigation-components}
\end{figure}

\newpage
\subsubsection{Layer Management Components}
The layer system provides a synoptic panel with a matrix-style interface for managing medical image layers:
\begin{itemize}
    \item \texttt{LayerMatrix}: coordinates overall layer display;
    \item \texttt{LayerMatrixRow}: manages individual layer controls;
    \item \texttt{LayerMatrixHeader}: provides column headers and global controls;
    \item \texttt{LayerTabs}: creates tabs on top of the panel to allow the user to switch between dataset layers and analysis results.
\end{itemize}

\begin{figure}[h]
    \centering
    \includegraphics[width=0.95\textwidth]{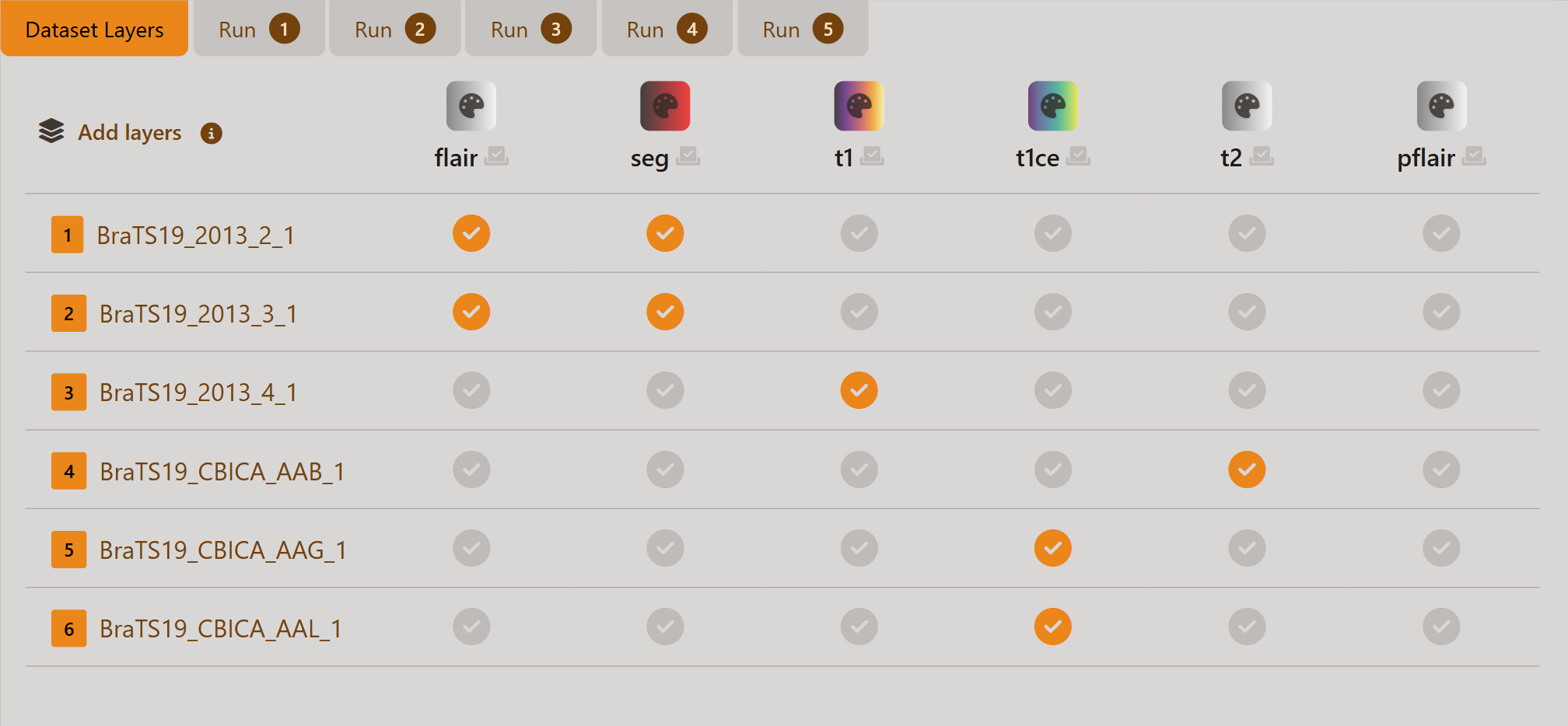}
    \caption{Dataset Layers interface showing multiple medical imaging datasets with their associated layers (flair, seg, t1, t1ce, t2, pflair). Orange checkmarks indicate visible layers, while gray checkmarks show available but hidden layers. Note the tabs at the top to switch between dataset layers and analysis results.}
    \label{fig:layers-interface}
\end{figure}

\subsubsection{Visualization Components}
The visualization components manage the layout of the viewers in the central panel:
\begin{itemize}
    \item \texttt{ViewersGrid}: manages the layout of multiple viewers;
    \item \texttt{ViewerWindow}: provides the container for individual viewers.
\end{itemize}

The core of our visualization system is built around NiiVue, a WebGL-based medical image viewer. The integration is implemented in a dedicated \texttt{NiivueViewer} component, which serves as a bridge between our application's layer management system and NiiVue's rendering capabilities. We implemented this component with careful consideration for the browser environment and resource management.

A key feature of our implementation is the sophisticated layer management system. The component maintains a reactive relationship with the application's Layer ViewModel, automatically responding to changes in layer selection and styling. This is achieved through a two-part system:

\begin{itemize}
    \item \textbf{Layer Loading:} when selected layers change, the component efficiently determines which layers need to be added or removed from the viewer. Instead of reloading all layers, it performs a differential update, removing only the layers that are no longer needed and adding only new ones;

    \item \textbf{Style Management:} the component implements a comprehensive color map and opacity management system that supports both predefined and custom color maps. For predefined maps (e.g. cyan, magenta, yellow...), these are registered during initialization. For custom color maps, the component can dynamically register new color maps using the layer's path as a unique identifier. As before, the updates of the layer's style are differentially updated.
\end{itemize}

The component also provides a public API for capturing screenshots, allowing users to save the current viewer state as PNG files named with a timestamp, making it reliable for clinical documentation purposes.

Concerning optimizations, the updates are optimized through debouncing, preventing performance issues when rapid changes occur, and resource cleanup is handled through Svelte's \texttt{onDestroy} lifecycle hook, ensuring that NiiVue's WebGL context and resources are properly disposed of when the component is unmounted. This prevents memory leaks and ensures proper application cleanup.

\begin{figure}[h]
    \centering
    \includegraphics[width=\textwidth]{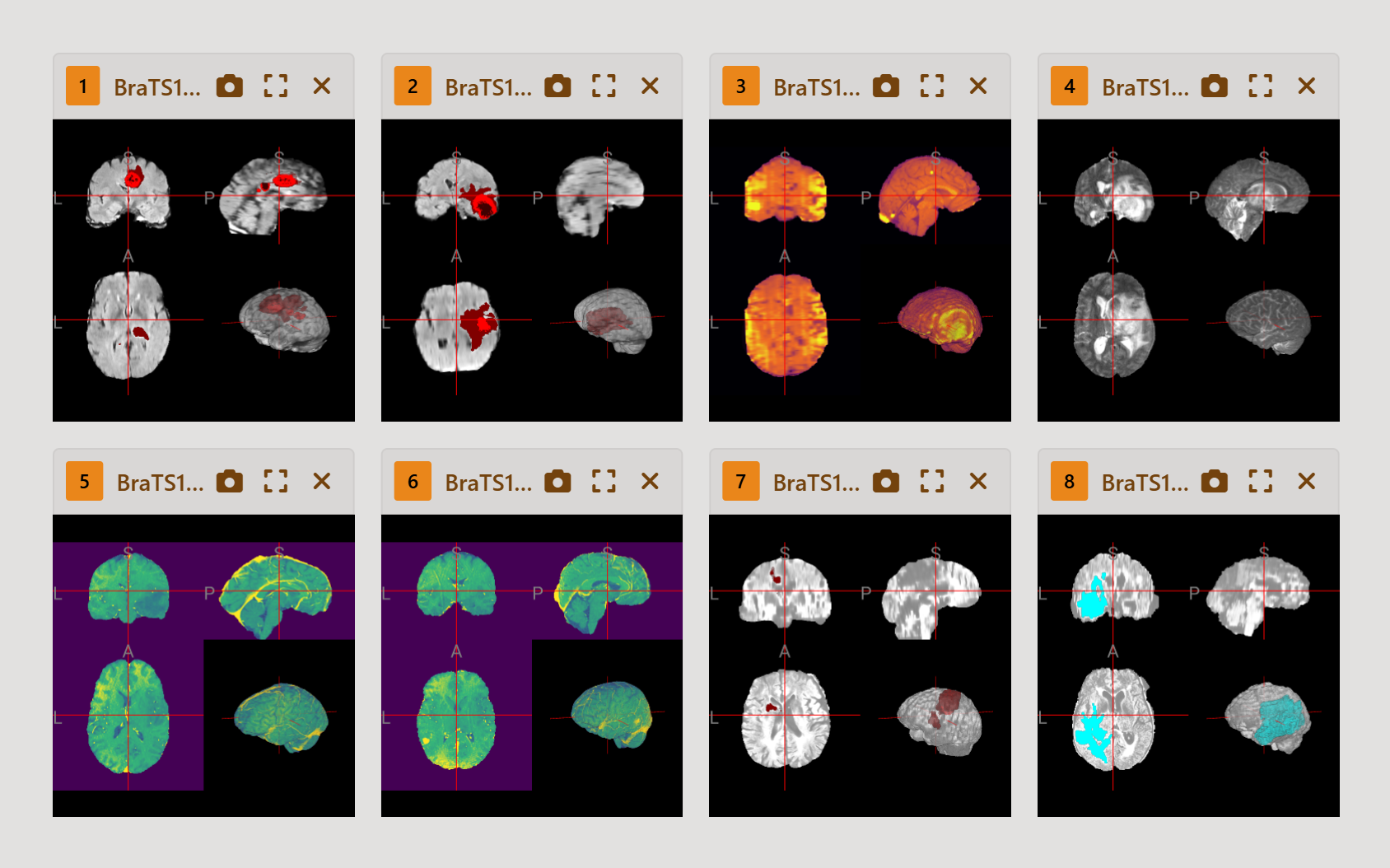}
    \caption{Multi-view medical image display interface showing eight concurrent viewer panels. Each panel displays orthogonal views (axial, sagittal, and coronal) of brain MRI data with different visualization modes and overlays. The interface supports synchronized navigation across viewers with consistent anatomical crosshairs and includes controls for image capture, fullscreen view, and panel closure.}
    \label{fig:image-viewers}
\end{figure}

% \mieke{Reading the chapter it could seem as if it is all standard and smooth. Were there any particular design issues that you felt were completely new to address and that required an original, novel solution to get it to work? It would be nice to highlight those in a discussion inserted close to where these decisions were made or as a general discussion at the end of the chapter for not breaking the flow of the presentation. In particular, I would think of how you had to `catch' the right elements from the ImgQL specification and turn them into active UI elements for layers and then connecting those to the viewer. Another part could address security issues and avoidance of leaking of information. You touched upon that, but it might be nice to expand on that and its importance for this domain of application. A third point is perhaps the workspace saving so that users can continue from previously obtained results. Not sure how novel this is, but it is very useful and perhaps in the web-based setting not so very common to find. }

% \vinc{I agree with Mieke's suggestion and would emphasize the difficulties encountered in the implementation as problems that are the objective of this chapter, or something along these lines, in the introductory bit of the chapter. Indeed saving state is not new at all and was sort-of-invented in web-based applications that require persistence to function and survive to browser reloads, but in the medical setting this brings in some innovative usage patterns! }

\newpage

\subsubsection{Analysis Components}
The analysis components provide tools for writing and executing spatial model checking operations:
\begin{itemize}
    \item \texttt{ScriptEditor} provides the ImgQL editing environment. The editor implementation includes custom syntax highlighting for ImgQL, integrated error reporting, and automatic script validation. Inside this component, we also handled the script execution, since it was fairly contained in terms of code length.
\end{itemize}

The automatic mapping from ImgQL scripts to UI layers presented unique challenges. We had to be careful to prevent the user from writing unsafe operations during the layers' importation, and find a way to make clear what are the outputs of the script and where they could be found.

The security challenge stemmed from the fundamental tension between Nielsen's usability heuristics \cite{nielsen1994usability} and medical data protection requirements. Allowing direct file path input - while satisfying Norman's visibility principle \cite{norman2014design} - risked exposing sensitive directory structures (information leakage) and potential path traversal attacks. Our solution combines automated script header generation based on UI state and a sandboxed script execution environment.

\newpage

The output management implementation required careful consideration of multiple UI design principles. Our initial prototypes revealed three potential approaches, each with distinct tradeoffs:

\begin{itemize}
    \item \textbf{Visualizing the outputs as additional columns in the layer management components:} while technically simple, this violated the \textit{simplicity} principle by creating visual clutter with more and more columns being added with each execution;
    \item \textbf{Visualizing the outputs in the browser section:} though logically organized, this broke \textit{consistency} with existing workflows and required duplicate control implementations;
    \item \textbf{Merged navigation components and layer management components:} created spatial conflicts that undermined \textit{progressive disclosure} of advanced features.
\end{itemize}

The final solution combines:
\begin{itemize}
    \item \textbf{Tabbed organization} mirroring dataset layer management (consistency principle);
    \item \textbf{Context-preserving expansion} that maintains spatial relationships (cognitive economy \cite{norman2014design});
    \item \textbf{Non-disruptive attention guidance} through subtle tab highlighting (blink duration 300ms, meeting WCAG 2.1 animation guidelines \cite{w3cWCAG}).
\end{itemize}

\begin{figure}[h]
    \centering
    \includegraphics[width=0.6\textwidth]{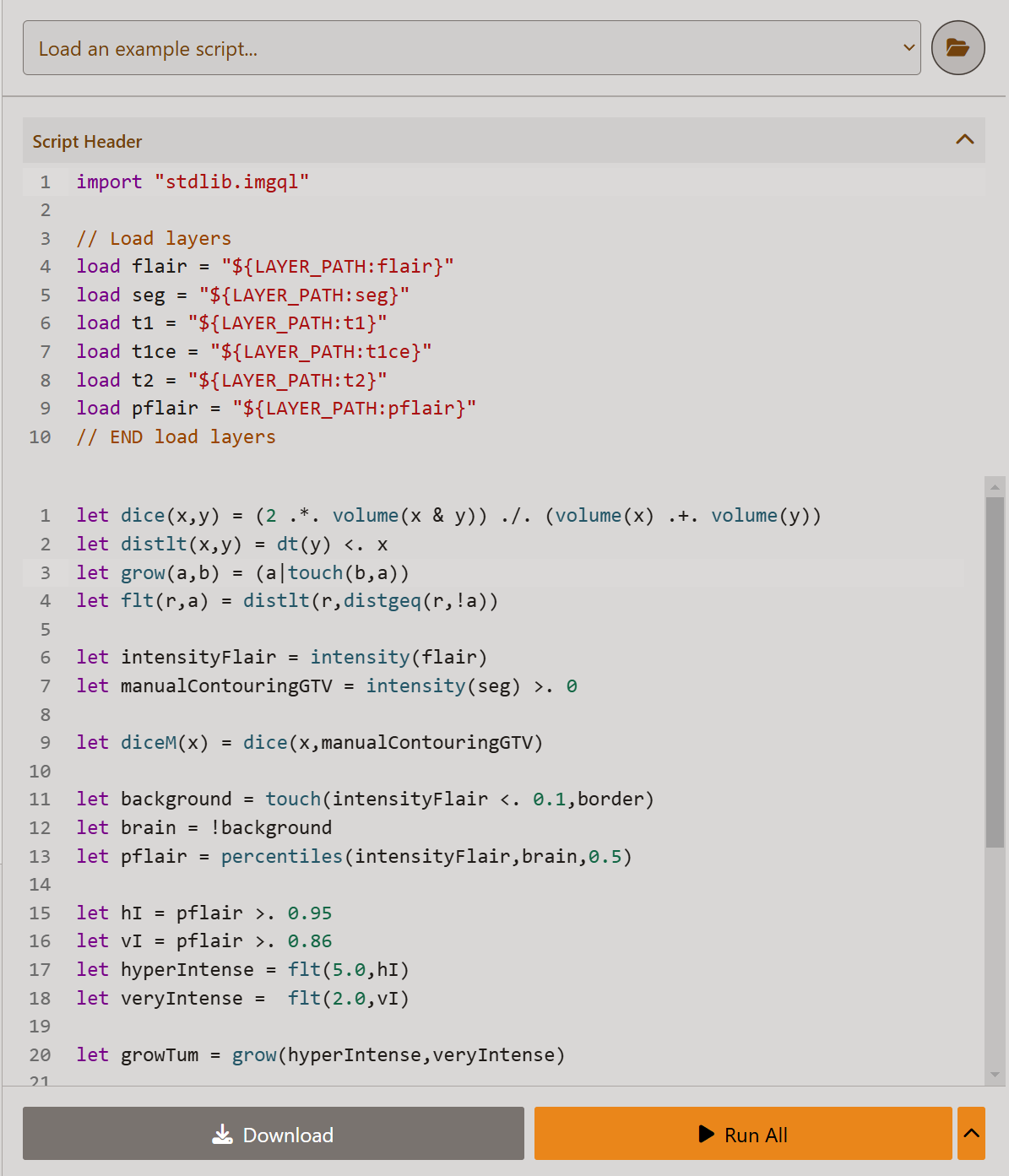}
    \caption{The sidepanel dedicated to the analysis components. It includes the script editor, a template loader (top), and execution controls (bottom).}
    \label{fig:script-editor}
\end{figure}

    \newpage

\section{Distribution}

The distribution strategy for VoxLogicA UI has been designed to accommodate both local installations and server deployments, providing flexibility in how users can access and utilize the application. This dual approach ensures the software can meet diverse usage scenarios, from individual researchers working on their local machines to medical institutions requiring centralized deployment.

\subsection{Local Installation}

For local installations, we have implemented several approaches to make the software accessible while maintaining the advantages of web technologies discussed in Chapter 3.

\subsubsection{Progressive Web App}

The user can simply download the application, run the docker file locally and access the application through the browser. To make the user experience more seamless, we have implemented the application as a Progressive Web App (PWA), leveraging modern web capabilities while maintaining the feel of a native application. This approach allows users to:

\begin{itemize}
    \item Access the application through any modern web browser;
    \item Install it locally through the browser's "Add to Home Screen" functionality.
\end{itemize}

\subsubsection{Desktop Application}
While currently not implemented, an Electron-based desktop application is currently being developed using Electron.js \cite{electronJS}, an open-source framework (maintained by OpenJS Foundation) that enables building desktop GUIs using web technologies (HTML, CSS, JavaScript) through Chromium and Node.js integration. This choice provides 2 main strategic advantages:

\begin{itemize}
    \item \textbf{Code Reuse:} almost all the web application's codebase can be directly reused, particularly the visualization components and UI logic developed in React.js;
    \item \textbf{Cross-Platform Deployment:} native packaging for Windows, macOS and Linux from single codebase, which can be easily distributed to users.
\end{itemize}

While Electron applications typically have larger binary sizes ($\sim$150MB) compared to native toolkits, this trade-off was deemed acceptable considering:
\begin{enumerate}
    \item Medical workstations' storage capacity;
    \item Rapid development cycle requirements;
    \item The team's existing web technology expertise.
\end{enumerate}

\newpage

\subsection{Deployment}

The deployment strategy has been implemented with a focus on containerization, ensuring consistent behavior across different environments while maintaining security and scalability.

% \mieke{The concept of containerisation is quite recent and it would deserve a few lines of introduction, motivation and references. This may be also useful in case domain experts in the medical world would be interested in reading your thesis. There are some recent references on wikipedia: https://en.wikipedia.org/wiki/Containerization\_(computing), but there may be more relevant ones. It would be good to add a reference to an accessible and reliable overview article that provides a discussion on these techniques. Based on that you could then add a few more lines on why it is particularly useful in the context of the topic of this thesis.} % ANTONIO: I've added an entire section in the background knowledge chapter.

The container configuration implements several key features:

\begin{itemize}
    \item \textbf{Multi-stage builds:} as briefly introduced before, we made sure to separate the build environment from the runtime environment to minimize the final image size;
    \item \textbf{Non-root user:} running the application as \texttt{node} user for enhanced security;
    \item \textbf{Volume management:} dedicated volumes for datasets, scripts, and workspaces;
    \item \textbf{Environment configuration:} flexible configuration through environment variables.
\end{itemize}

    % % Evaluation
    \chapter{Evaluation}

The development of any medical system demands rigorous validation to bridge the gap between technical capability and clinical applicability. In this chapter, we transition from the VoxLogicA UI's design to its practical assessment through a series of evaluations with domain experts.

We have performed a series of rigorous user tests to assess the usability of the VoxLogicA UI both with researchers and medical students, followed by more informal interviews and presentations with expert clinicians and researchers to start gathering attention and feedback from a broader audience.

\section{User Tests}

Usability in medical tools is critical, as it directly impacts workflow efficiency, cognitive load, and the adoption of new technologies in clinical practice. Generally usability tests involve a limited number of users (typically five of them) to uncover the majority of the most common problems in the GUI \cite{nielsen2000howManyTestUsers}.

In this work, we performed two qualitative studies: a first study conducted with a group of users belonging to the research area and a second study conducted with a group of medical physicists students. Each study involves questions on the GUI design choices we described in the previous chapters, and then takes a separate direction for questions dedicated to the specific workflow of the users. The test material, the post-study questionnaire and the results of the post-study questionnaire are available in \cite{voxlogicAUserTests}, both in English and in Italian (the latter language was used to conduct the experiments). Below we present the study methodology, the partecipant selection, the testing procedures for the two studies, and a discussion of the results.

\subsection{Methodology}

The methodology has been the same for both studies, structured into 3 steps:

\begin{enumerate}
    \item \textbf{Training}: since most of our participants had varying degrees of familiarity with the concepts involved (e.g. VoxLogicA, layers, similarity indexes etc.), ranging from no prior knowledge to some expertise, we trained all of them by showing the GUI and its features with the aid of a short video tutorial produced for the purpose. The video tutorial is available on YouTube in italian language \cite{voxlogicauiTutorial};
    \item \textbf{Task performance}: participants are then asked to perform selected representative tasks in a realistic scenario. For each task, participants provided an answer or performed an action on the GUI. The interaction with the GUI has been conducted under the supervision of a test moderator who guides and observes the participants during the execution of the tasks, annotating the answers and the participants' behaviour (e.g. hesitations, feedback, comments, and self-confidence). At the beginning of the test, we also allowed participants to freely explore the UI for 1 minute to get a first impression and ask questions to the moderator, as a bridge to the training video;
    \item \textbf{Questionnaire}: participants were asked to fill out a post-study anonymous questionnaire to elicit details and feedback. We have structured it in 3 sections:
    \begin{enumerate}
        \item Information about the partecipant's demographic data (e.g. age, education, employment, work area, level of knowledge of VoxLogicA and medical image analysis). This information has been used to check whether some demographic variables determine differences in the test results, thus potentially acting as moderator variables;
        \item Questions about the GUI design choices and the features used during the test, with a focus on the clarity, completeness, and easiness of navigation of the GUI elements. Some of the questions required an evaluation on a Likert scale from 1 (not at all clear) to 5 (extremely clear), other prompted open-ended answers;
        \item A SUS (System Usability Scale) questionnaire to assess user satisfaction and perceived usability of the GUI \cite{sus}. Our SUS questionnaire contained 10 general statements concerning the use of the GUI prototype, evaluated on a Likert scale from 1 (strongly disagree) to 5 (strongly agree). Odd numbered statements expressed positive evaluations about the GUI prototype, while statements with an even number expressed negative evaluations.
    \end{enumerate}
\end{enumerate}

\subsection{Participant selection}

As anticipated, the studies have been conducted on small groups of users (5 participants per study):

\begin{itemize}
    \item \textbf{First study}: participants belonged to the research area: they were computer science researchers with a background in software verification and validation. Among them, only one knew about VoxLogicA, and two of them had some prior knowledge of medical image analysis;
    \item \textbf{Second study}: participants belonged to the clinical area: they were medical physics students with a background in medical image analysis. None of them knew about VoxLogicA, but all of them had a good knowledge of medical image analysis.
\end{itemize}

\subsection{Testing procedures}

The 2 studies were conducted on the same GUI version, and an almost identical set of tasks: for both studies we tested the user interaction with all the visual components we created, but in the first study we put more emphasis on the script editing and execution, while in the second we focused on the result visualization and analysis.

More specifically, a summary of each macrotask we tested for each study follows:

\begin{itemize}
    \item \textbf{First study}:
    \begin{itemize}
        \item Workspace creation and selection;
        \item Dataset navigation, case selection, and search functionality;
        \item Layers management and visibility of dataset cases;
        \item Script selection/execution and result interpretation capabilities (i.e. identify and understand the DICE index, view and understand the output layers);
        \item Script editing capabilities and layer customization features;
        \item UI customization and view adjustment features.
    \end{itemize}
    \item \textbf{Second study}:
    \begin{itemize}
        \item Workspace creation and selection;
        \item Dataset navigation, case selection, and search functionality;
        \item Layers management and visibility of dataset cases;
        \item Script selection/execution and result interpretation capabilities (i.e. identify and understand the DICE index, view and understand the output layers);
        \item Case filtering based on similarity index;
        \item Detailed result inspection and comparison features (i.e. compare manual vs automatic tumor segmentation);
        \item Session management and UI customization features.
    \end{itemize}
\end{itemize}

This was done to explore more in depth the different workflows each user would have, depending on their needs and background.

\subsection{Results analysis}

To have a measure of the perceived comprehensibility and usability of the several GUI sections and features tested, we first analyzed the responses from the post-study questionnaire. For each question to be answered with a Likert scale, we computed the average score for each section/feature tested, while we collected feedback and suggestions through open-ended questions.

\newpage

Then, we analyzed the data collected by the test moderator during the tasks performance step, to get a measure of the tasks' success rate and some qualitative evaluation gathered by the observations, identifying the most common issues and difficulties encountered by the users. The tasks' success rate is computed as the sum of the tasks completed successfully (with tasks completed with some issue each weighting 0.5), divided by the total number of completion attempts (i.e. the product of the number of tasks and the number of participants).

Finally, to further enrich the evaluation, we incorporated the results obtained from the SUS questionnaire, which yields a single score on a scale of 0-100 representing a measure of the perceived usability of the GUI and user satisfaction. A score of 71 or higher is considered good \cite{susAdjectiveRatingScale}.

\subsubsection{Post-study questionnaire}

The post-study questionnaire revealed generally positive evaluations across all GUI components, with specific areas for improvement:

\begin{itemize}
    \item \textbf{General interface clarity} scored 4.4/5 overall. While 90\% of participants rated it 4-5/5, one user suggested differentiating sections with different background colors/borders to enhance the visual separation between sections;

    \item \textbf{Workspace management} received exceptional ratings (avg. 4.9/5), with perfect scores from 9/10 participants and no issues reported;

    \item \textbf{Dataset navigation} scored well (avg. 4.6/5), though some participants suggested improved visual differentiation between datasets using color-coding or spacing, and show additional information about the dataset (e.g. information about the patients or the date of the studies);

    \item \textbf{Dataset cases management} received strong ratings (avg. 4.6/5). Participants particularly appreciated the case selection mechanism (7/10 rated 5/5), though three suggested improvements: visual differentiation through spacing/colors (P02, P07), and a database-style interface for medical metadata (P07);

    \item \textbf{Search functionality} showed very high scores (avg. 4.7/5), with two participants noting empty search results should be explicitly indicated and one requesting a better highlighting of the results (to differentiate them from the rest of the items);

    \item \textbf{Layer management} received good scores for individual layer visiblity toggling (avg. 4.4/5) and for all layers visibility toggling (avg. 4.5/5), but some participants noted the need for a more prominent visual indicator for the all layers visibility toggling;

    \item \textbf{Script execution} showed good scores for preset selection (avg. 4.3/5) and code editor (avg. 4.4/5). A partecipant expressed concern about auto-save (possibly suggesting version control introduction), while another would have liked a better color palette;

    \newpage

    \item \textbf{Image viewer} functionality received a very high score for basic visualization (avg. 4.8/5), with interface controls scoring slightly lower (avg. 4.7/5). One participant reported scrollbars occasionally obscuring window controls;

    \item \textbf{UI customization} features like panel resizing scored well (avg. 4.4/5), and the icons are generally well perceived (avg. 4.3/5), but the day/night mode icon caused confusion (2/10 participants mistook it for settings). Participants suggested alternative sun symbols or toggle switches.
\end{itemize}

\subsubsection{Tasks performance}

Concerning the first study, there was a total of 34 tasks per participant. We obtained a remarkable success rate of 94\%: this resulted from no failed tasks and 19 partial completions (0.5 weight each) due to some interface challenges:

\begin{itemize}
    \item \textbf{Visual clarity issues}: small DICE value text in RUN panel and layers rendering order difference in between cases (which happens when opening layers for the cases in different orders);
    \item \textbf{Color management}: some users found the colormap options overwhelming;
    \item \textbf{Code editor in dark mode}: 80\% of participants reported legibility problems in the code editor when dark mode was enabled.
\end{itemize}

Other problems not related to the interface design but that some participants encountered were one with the individual viewers' controls not being visible when the viewers section showed a scrollbar (even though this was more of a bug than a usability issue), and one where one participant struggled to understand the ImgQL syntax. We still counted them as partial completions.

Concerning the second study, there was a total of 37 tasks per participant. We obtained another remarkable success rate of 96\%: this resulted from no failed tasks and 13 partial completion (0.5 weight each) due to some interface challenges:

\begin{itemize}
    \item \textbf{Code editor panel management}: 4 participants struggled to collapse the code editor section, searching unsuccessfully for a dedicated close button;
    \item \textbf{Workspace switching}: 1 user changed workspace through the top-left logo instead of the dedicated workspace selector;
    \item \textbf{Layer management}: 2 participants had difficulty closing opened layers, while one user confused output layers with dataset layers when searching for specific segmentation results;
    \item \textbf{Script execution workflow}: some users experienced initial confusion locating both the script execution panel (3 participants) and preset selection interface (2 participants);
    \item \textbf{Filter interface}: 2 participants expressed uncertainty about filter application mechanics, with one confusing filtered case counts with opened case counts;
    \item \textbf{UI elements}: the light/dark mode toggle icon continued to cause confusion (2 users), mirroring findings from the first study;
    \item \textbf{Bulk operations}: 3 users spontaneously attempted to open all cases in a dataset simultaneously, suggesting unmet needs for batch processing.
\end{itemize}

A user encountered another bug, this time related to the color customization, where the opacity control bar occasionally became hidden during custom color selection. We counted this as a partial completion.

\subsubsection{SUS questionnaire}

In the SUS questionnaire we achieved an excellent mean score of 90.0/100, with only a single participant scoring 70.0/100 \cite{saurometerSUSInterpretation} and everyone else scoring above 85.0/100 (which is considered excellent).

\begin{figure}[htbp]
    \centering
    \includegraphics[width=\textwidth]{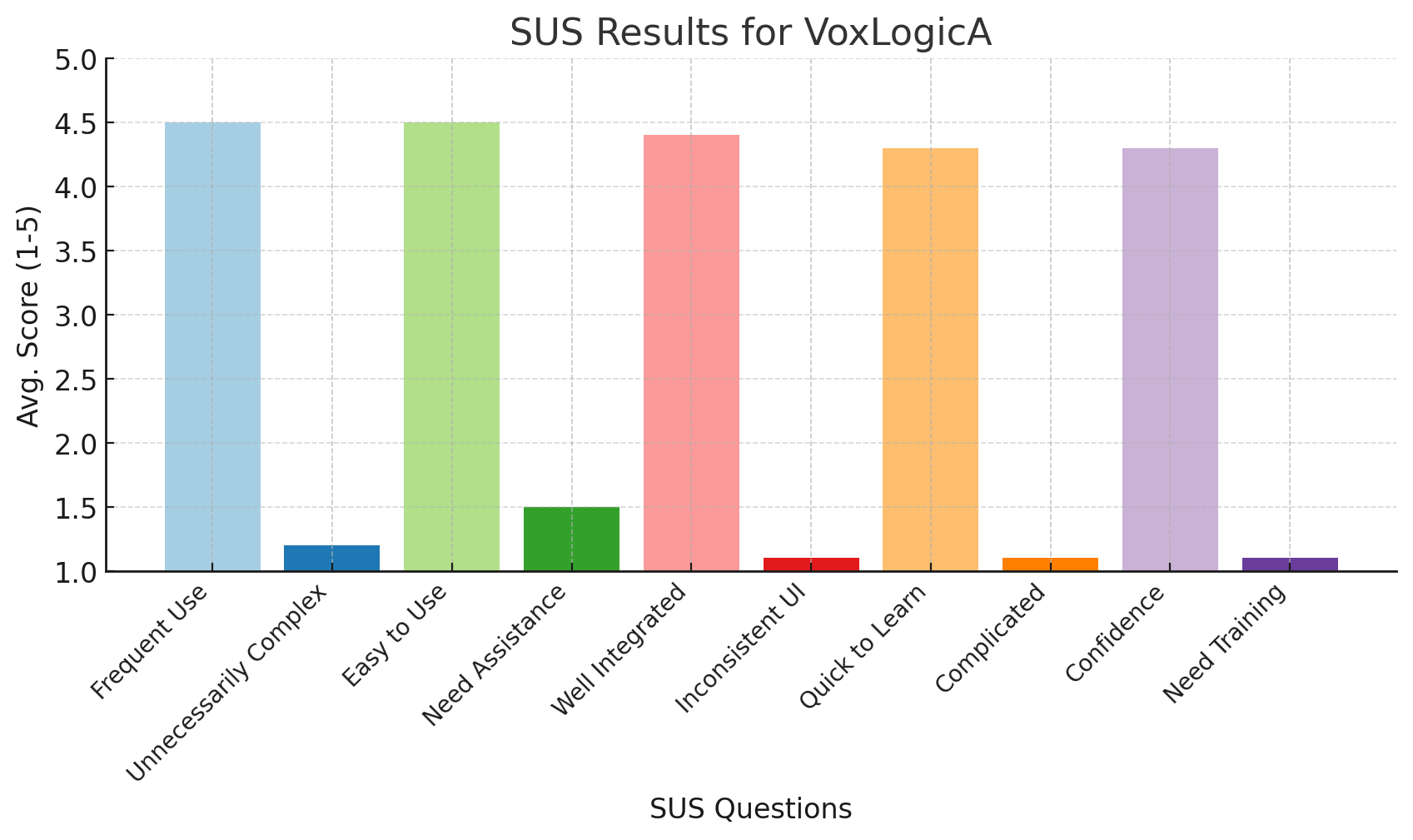}
    \caption{System Usability Scale (SUS) detailed results showing average scores for each question category. Higher scores on positive statements (e.g., Frequent Use, Easy to Use) and lower scores on negative statements (e.g., Unnecessarily Complex, Need Training) indicate better usability.}
    \label{fig:sus_scores}
\end{figure}

\subsubsection{Discussion}

The evaluation results demonstrate the prototype VoxLogicA UI appears to successfully balance technical capability with clinical usability, while revealing valuable insights for refinement. Three key findings emerge from our mixed-methods analysis:

\begin{enumerate}
    \item \textbf{High usability baseline}: the exceptional SUS score (90.0) and task success rates (94-96\%) surpass established thresholds for software acceptability, indicating strong foundational usability. This aligns with Nielsen's efficiency of use principle \cite{nielsenUsabilityEngineering}, particularly evidenced in workspace management (4.9/5) and basic visualization (4.8/5);
    \item \textbf{Context-sensitive design requirements}: the divergence in problematic usage patterns between user groups underscores the importance of role-specific customization. Researchers emphasized script editing improvements (color schemes, version control), while medical users prioritized batch operations and metadata access - a dichotomy reflecting the tool's dual use for development and clinical analysis;
    \item \textbf{Visual perception challenges}: recurring issues with icon semantics (light/dark mode toggle) and layer differentiation echo Gestalt principles of perceptual organization \cite{gestaltLawsPerceptualOrganization}. The 80\% error rate in dark mode legibility particularly violates Nielsen's visibility principle \cite{nielsenUsabilityEngineering}, suggesting need for adaptive contrast ratios.
\end{enumerate}

Notably, participants spontaneously attempted bulk operations (3/10), revealing an unmet need that is common in research workflows. This emergent requirement, coupled with high satisfaction in core functionality, suggests our interface successfully supports essential tasks while exposing opportunities for advanced features.

    \section{Public Exposure}
Following successful prototype validation through controlled qualitative user tests, we pursued broader exposure through multiple channels to assess real-world viability and gather diverse stakeholder perspectives.

\subsection{Open Source Release}
The VoxLogicA UI project entered public beta in December 2024 with the release of:
\begin{itemize}
    \item Public GitHub repository: \\ \url{https://github.com/VoxLogicA-Project/VoxLogicA-UI}
    \item Official project website: \\ \url{https://voxlogica-project.github.io/VoxLogicA-UI/}
\end{itemize}

The website functions as a demonstration platform and project showcase, featuring annotated screenshots of key functionalities and a tutorial video to help users get started, while providing direct access to the development repository.

\subsection{Technical Collaborations}
We established contact with the Niivue development team \cite{niivueMeeting2025}, receiving positive feedback on our interface design. This dialogue opened the door to potential collaborations, including joint development on a general purpose medical image viewer (with plugins for VoxLogicA analysis), and several optimizations for multiple cases visualization as we implemented in our work (since Niivue is currently showing performance issues in these cases).

\subsection{Clinical Engagement}
We presented our work at Bari Hospital through a couple of video meetings in January 2025, generating valuable insights from medical professionals.

A \textbf{general practitioner} and a \textbf{doctor specialising in radiodiagnostics}, Dr. Michele Catalano, praised the minimalistic visual organization and quantitative analysis capabilities, particularly appreciating:
\begin{itemize}
    \item Immediate DICE score feedback as a aid for treatment planning;
    \item Intuitive case comparison workflows;
    \item Color-coded differentiation for the different layers.
\end{itemize}

In particular, Dr. Michele Catalano provided detailed technical feedback:

\begin{itemize}
    \item \textbf{Visual Design:}
    \begin{itemize}
        \item Emphasized the clinical necessity of dark mode: "High-brightness medical monitors (e.g., Eizo RX660) used in radiology make light themes visually fatiguing during prolonged reporting sessions";
        \item Suggested exploring even darker interface options for extended diagnostic workflows.
    \end{itemize}
    
    \item \textbf{Viewer Enhancements:}
    \begin{itemize}
        \item \textbf{Axis Reorientation:} critical for analyzing misaligned scans common in trauma cases;
        \item \textbf{Window Synchronization:} simultaneous viewing of correlated image planes with linked scrolling;
        \item \textbf{Quantitative Tools:} demand lesion measurement tools and ROI (Region of Interest) density analysis, particularly for Digital Volume Imaging (DVI) applications.
    \end{itemize}
    
    \item \textbf{Segmentation Utility:}
    \begin{itemize}
        \item Validated the clinical relevance but noted: "The DICE metric requires clearer explanation for non-technical users";
    \end{itemize}
\end{itemize}

This specialist feedback aligns with established radiology workstation requirements \cite{fdaGuidanceDisplayDevicesDiagnosticRadiology} while highlighting opportunities for domain-specific optimizations.

\subsection{Academic Reception}
Finally, we had a chance to present a demo of our work to a group of medical students at the University of Bari's School of Medicine and Surgery.

With 10 minutes of explanation on VoxLogicA, the students were already able to understand the interface and the main features: they particularly appreciated the minimalistic design and the quantitative analysis capabilities, and 2 students expressed their highest interest saying that they would be willing to use VoxLogicA UI already as it is in the current state, since the viewers they are currently employing are not very user-friendly.

    % % Discussion and Conclusion
    \chapter{Conclusion}

In this work we have aimed at making a significant step forward in making spatial model checking accessible to medical researchers and clinicians. 

By bridging the gap between VoxLogicA's powerful analytical capabilities and modern human-computer interaction principles, we have created a tool that combines formal methods rigor with clinical usability.

Our web-based approach has successfully addressed the dual challenge of healthcare IT constraints and cross-platform compatibility. The architectural choices - particularly the MVVM pattern implementation with Svelte 5's reactivity system and Docker-based deployment - have proven effective in creating a maintainable, scalable solution. The evaluation results (94-96\% task success rates, SUS score of 90) are very promising and seem to validate our design decisions, showing that complex medical image analysis can be made accessible without sacrificing technical capability.

\section{Impact and Limitations}
The project demonstrates three key advancements in medical imaging tools:
\begin{itemize}
    \item \textbf{Democratized Access}: web deployment eliminates traditional software installation barriers while maintaining desktop-like capabilities through PWA technology;
    \item \textbf{Workflow Integration}: the workspace system with atomic updates and differential storage enables longitudinal analysis sessions mirroring clinical workflows;
    \item \textbf{Visual Model Checking}: leveraging the successful points of the predecessor of this work - VoxLogicA Explorer - the tight integration between ImgQL scripts and layered visualization creates a novel feedback loop for spatial property validation.
\end{itemize}

However, limitations remain. While our Docker solution addresses deployment challenges, some clinical environments may prefer traditional installers - an area for potential Electron-based expansion. Additionally, data accessibility remains constrained by requirements for Docker volume mounts and specific directory structures, creating usability barriers for non-technical practitioners. Furthermore, while VoxLogicA supports common image formats like JPEG and PNG, our current implementation's focus on NIfTI files limits its applicability to broader medical imaging use cases, though this primarily reflects the current GUI implementation rather than fundamental technical constraints, and could be addressed through future interface enhancements.

% \mieke{The last point is the current state of the GUI, but there are no serious technical problems to overcome these limitations in the future I suppose? Would be good to mention.}

\section{Future Directions}
In addition to the immediate fixes and resolution of the problems identified in the evaluation, several promising avenues for further development emerge:
\begin{enumerate}
    \item \textbf{Clinical Data Exploration}: responding to medical community interest, developing a standalone DICOM database explorer/viewer mode would make spatial analysis accessible without requiring ImgQL expertise, while preserving VoxLogicA integration as an advanced plugin;

    \item \textbf{VoxLogicA 2}: building on the work in \cite{hybridAIvoxlogica2024}, integration with the next version of VoxLogicA, VoxLogicA 2, should emphasize human-AI collaboration - maintaining clinician oversight for all diagnostic decisions while leveraging automated pattern detection;

    \item \textbf{Automated Pattern Suggestions}: developing AI-assisted recommendation systems that analyze image features (texture, morphology, intensity distributions) to propose relevant ImgQL spatial logic patterns, lowering the barrier to entry for clinicians without formal methods training while maintaining expert oversight through interactive validation interfaces;

    \item \textbf{Hospital System Integration}: direct PACS connectivity would enable real-time analysis of clinical imaging workflows while maintaining strict HIPAA compliance through our differential storage architecture.
\end{enumerate}

This work ultimately shows that formal methods need not to remain confined to verification laboratories. By embracing modern web technologies and user-centered design, we can bring mathematical rigor to clinical practice - not as an abstract concept, but as a tangible tool in the diagnosis and treatment planning of neurological disorders. The positive reception from both computer scientists and medical professionals suggests we have taken an important step toward this vision.

    % References
    \bibliographystyle{plain}
    \bibliography{main.bib}
    
    % Acknowledgements
    \chapter*{Acknowledgements}
\thispagestyle{empty}

Innanzitutto, vorrei esprimere la mia più profonda gratitudine ai miei supervisori Vincenzo Ciancia, Fabio Gadducci e Mieke Massink per la loro guida esperta, pazienza e preziosi consigli durante questo percorso di ricerca e sviluppo. Un ringraziamento speciale va a Giovanna Broccia per il suo contributo fondamentale nella progettazione e conduzione degli user test, ambito per me inizialmente ignoto.

Alla mia famiglia: il vostro supporto incondizionato è stato la roccia sulla quale ho costruito ogni traguardo. I sacrifici fatti per la mia formazione risuonano in ogni dimostrazione fatta e in ogni riga di codice scritta.

A Daniele, che merita il titolo di Generale del Lavoro di Tesi, i cui messaggi quotidiani "E anche oggi si insegue il nostro sogno" sono stati il metronomo instancabile che ha scandito il ritmo di questo lavoro. Alessio, grazie per essere il mio sensei della sicurezza: possano i nostri sistemi rimanere sempre inviolati. A Francesco, l'eroe HCI (mai così necessario, mai così immeritato) dietro un numero non meglio specificato di consulenze sulla UI e UX del progetto. Graziano, il manager indiscusso, maestro di social skills: grazie per avermi fatto sentire accolto per la prima volta a Pisa ed essere stato presente nei momenti bui. A Edoardo, mio complice in ogni battuta a sfondo scientifico, nonché co-inventore del razzo ad acqua: grazie per avermi ricordato il bello di fare ricerca. Simone, grazie per essere stato il mio empatico e instancabile nutrizionista: possano i nostri pasti essere sempre bilanciati come le tue spiegazioni. Giulio, Lorenzo: grazie per avermi mantenuto sulla "retta via" durante ogni esame alla magistrale fatto insieme. A Bianca, mia (unica) complice negli indovinelli su YouTube. A Desi, per avermi re-insegnato come si pusha (no, non i commit, in palestra).

Un ringraziamento speciale a mio cugino Michele per avermi messo in contatto con professionisti medici che attualmente lavorano nella regione Puglia, il cui feedback specializzato è stato di fondamentale importanza per la validazione del sistema. Un grazie particolare anche a tutti i partecipanti anonimi che hanno generosamente dedicato il loro tempo ai test di usabilità dell'interfaccia, trasformando ogni sessione di test in un'opportunità di miglioramento.

Ringrazio infine amici e colleghi che mi hanno dedicato tempo in questi anni, architetti inconsapevoli dell'edificio intellettuale che oggi mi porto dentro.

\begin{flushright}
\textit{Antonio Strippoli}
\end{flushright}

\end{document}